\documentclass[]{aastex631}
\usepackage{booktabs}
\usepackage{xcolor}
\usepackage{float}
\usepackage{footnote}
\usepackage{amsmath}
\usepackage{amssymb}
\usepackage{bbm}
\usepackage{dsfont}
\usepackage{graphicx}
\usepackage{subfigure}
\usepackage{float}
\usepackage{multirow}
\usepackage{array}
\newcolumntype{b}{!{\vrule width 1.5pt}}

\newcommand{\obsRV}{\phi_{\textrm{obs}}}
\newcommand{\photonRV}{\phi_{\textrm{pho}}}
\newcommand{\readNoiseRV}{\phi_{\textrm{read}}}
\newcommand{\darkNoiseRV}{\phi_{\textrm{dark}}}
\newcommand{\quantNoiseRV}{\phi_{\textrm{quant}}}

\newcommand{\bighline}
{\noalign{\global\arrayrulewidth1.5pt}\hline\noalign{\global\arrayrulewidth0.4pt}}

\shorttitle{AstroSURE:‌ Learning to Remove Noise Without Ground Truth}
\shortauthors{Vaheb et al.}
\graphicspath{{./}{figures/}}

\begin{document}

\title{AstroSURE: Learning to Remove Noise from Astronomical Images Without Ground Truth Data}

\author[0009-0008-2192-089X]{Omid Vaheb}
\affiliation{Edward S.~Rogers Sr.~Department of Electrical and Computer Engineering, University of Toronto, 10 King's College Road, Toronto, ON M5S 3G4, Canada}

\author[0000-0003-2239-7988]{Sébastien Fabbro}
\affiliation{National Research Council Herzberg Astronomy an Astrophysics Research Centre, 5071 West Saanich Road, Victoria, BC V9E 2E7, Canada}
\affiliation{Department of Computer Science, , University of British Columbia, Vancouver, BC, V6T 1Z4, Canada}

\author{Stark C.~Draper}
\affiliation{Edward S.~Rogers Sr.~Department of Electrical and Computer Engineering, University of Toronto, 10 King's College Road, Toronto, ON M5S 3G4, Canada}



\begin{abstract}

In astronomical imaging, the low photon count of exposures necessitates extensive post-processing steps including contamination removal and denoising. This paper evaluates deep-learning denoising methods that can be trained without clean ground-truth images and assesses their utility for detection-oriented analysis of astronomical data. 
We adapt and compare Noise2Noise, Stein’s Unbiased Risk Estimator, and blind-spot-based methods using synthetic data and real observations from the Hubble Space Telescope (HST) and the Canada-France-Hawaii Telescope (CFHT). Performance is evaluated using object-detection metrics, including correct detection rate and false alarm rate, together with image-based metrics and pixel-distribution diagnostics. The results show that these methods can improve faint-source detectability relative to the original noisy images, with encouraging gains on HST data after domain-consistent initialization, while transfer to CFHT data is more limited, highlighting the importance of instrument/domain similarity for unsupervised adaptation.


\end{abstract}



\section{Introduction} \label{sec:intro}
Astronomical imaging pushes the boundaries of detection, often dealing with very low photon counts from faint and distant sources. Individual images, whether single exposures or the product of a combination of exposures, are inevitably contaminated by various noise sources, including photon shot noise and detector readout noise. This noise can obscure faint structures and limit the precision of scientific measurements like photometry and object detection \citep{see_in_dark, classical_astro_denoising}. Enhancing the faintest scientifically relevant signals is therefore an important goal. 

When $N>1$ exposures of the same field are available, a standard approach is to combine them ("stacking") to improve the signal-to-noise ratio (SNR), theoretically by a factor of $\sqrt{N}$ for ideal averaging of similar exposures \citep{swarp}. However, achieving this optimal gain requires careful treatment of outliers like cosmic rays or detector artifacts. Robust methods like median stacking effectively reject outliers but may not provide the best SNR improvement, especially for low $N$. Conversely, methods like weighted averaging can approach the theoretical SNR gain but are highly sensitive to unrejected outliers, and automated outlier rejection algorithms themselves may fail or misidentify faint sources when $N$ is small \citep{swarp, hst_processing}.

Furthermore, multi-exposure datasets are not always available due to observational constraints, time-domain studies, or the nature of specific calibration frames. Therefore, methods capable of improving the quality of single astronomical images are useful, both for analyzing individual exposures directly and potentially for enhancing the final product of a stacking pipeline. The primary aim is to reduce noise and enhance the visibility of the faintest structures possible within any given image \citep{classical_astro_denoising}. 

Additionally, for precise quantitative measurements (e.g., galaxy shape or photometry), modeling the characteristics, such as Point Spread Function and noise properties, of individual exposures can be more tractable than modeling those of a complex, stacked image of resampled individual exposures. Indeed, fitting models simultaneously to multiple individual exposures has been shown to yield robust measurements, bypassing the difficulties associated with characterizing the final stacked product \citep{galfitm}. This further motivates techniques that can operate effectively on individual exposures to either denoise the input image for a detection objective or effectively mark the contaminant in a weight or mask image for traditional source property measurements.

Deep learning (DL) methods have shown considerable success in general image denoising tasks \citep{dncnn}. However, applying standard supervised DL techniques to astronomical data faces an obstacle, as clean, noise-free reference images are unavailable for astronomical observations. While some studies have used heavily stacked images as approximations of reference data \citep{astro_unet, astro_unet_fake}, this approach has limitations, can be computationally expensive, and is not universally applicable. Recent advancements in unsupervised and self-supervised learning offer alternatives, enabling networks to be trained directly on noisy observational data \citep{n2n, noise2void, noise2self, self2self}. While some applications exist in specific astronomical contexts \citep{solar_denoising, unsupervised_astro_denoising}, a systematic evaluation of state-of-the-art techniques for general astronomical image denoising, focusing on scientifically relevant metrics, is needed.


This paper introduces AstroSURE, a practical framework for evaluating denoising methods that can be trained without clean astronomical targets. It operates on two settings that are often relevant in practice: paired noisy observations, handled with Noise2Noise, and single noisy images, handled with SURE. Using synthetic data and exposures from HST and CFHT, we assess whether these methods improve faint-source detection in individual exposures. Our emphasis is therefore on detection-oriented performance, quantified with correct detection rate (CDR) and false alarm rate (FAR), rather than on precision photometry, morphology, or other PSF-sensitive measurements. In this sense, the main contribution of AstroSURE is a practical comparison of N2N, SURE, and blind-spot methods for astronomical image denoising, including an examination of domain transfer from simulations to real space- and ground-based data and of the role of domain-consistent initialization in unsupervised training. 

In \autoref{sec:background}, we review learning-based denoising techniques in the context of the requirements of astronomical imaging. In \autoref{sec:data}, we describe the data used, including synthetic data generated by {\em Galsim} library and real data from HST and the CFHT. We test our approach using realistic synthetic data and real observations, providing a practical framework for improving astronomical data quality. We describe our proposed AstroSURE in \autoref{sec:method}, followed by the results and experiments in \autoref{sec:experiments}. Finally, in \autoref{sec:conclusion}, we conclude the work and suggest the next steps.

\section{Background} \label{sec:background}

Images in astronomy are fundamentally photon-limited, meaning noise often obscures the faint signals essential for scientific discovery. While traditional denoising techniques like Gaussian smoothing or BM3D exist \citep{classical_astro_denoising}, their effectiveness can be limited by the complex nature of astronomical noise and the need to preserve faint source morphology for reliable detection and subsequent analysis. Deep learning, particularly using Convolutional Neural Networks (CNNs) are powerful alternative, capable of learning intricate patterns from data \citep{dncnn, cnn1, cnn2}.

Standard deep learning approaches in astronomy denoising are typically "supervised," requiring pairs of noisy images and corresponding noise-free "ground truth" targets for training. This presents a major hurdle where such clean reference images are unobtainable, basically every scenario in observable astronomy. Using heavily stacked images as proxies \citep{denoisingdeepsky, astro_unet, astro_unet_fake} is a common workaround but suffers from practical limitations: requiring significant observational data, potential introduction of artifacts during stacking, and models often lacking generalizability across different instruments or observing conditions \citep{noise2astro, autoencoder_astro, unsupervised_astro_denoising}. These issues necessitate denoising methods that can learn effectively without clean reference data.

Alternatives arise from unsupervised and self-supervised learning. Noise2Noise (N2N) demonstrated that networks can be trained using only pairs of noisy images of the same scene, provided their noise is independent \citep{n2n}. While effective, N2N's requirement for suitable image pairs is not always met in practice. Other self-supervised methods operate on single noisy images, such as blind-spot techniques (e.g., Noise2Void, Noise2Self) \citep{noise2void, noise2self} or approaches based on Stein's Unbiased Risk Estimator (SURE) \citep{SURE}. SURE, adaptable to mixed Poisson-Gaussian noise \citep{poisson_gaussian_sure, sure_deep, esure}, allows estimating the denoising error from the noisy image alone, given knowledge of the noise parameters.

These less supervised methods have seen initial application in astronomy. N2N has been used for solar data \citep{solar_denoising} and limited simulations \citep{noise2astro}. Self-supervised methods like Self2Self \citep{self2self, unsupervised_astro_denoising} and Noisier2Noise \citep{noiser2noise, autoencoder_astro} have also been explored. However, blind-spot networks tend to suppress the underlying signal and introduces noticeable grid-like artifacts \citep{self2self}. Furthermore, methods relying on single-image training combined with heuristic post-processing, such as using arbitrary thresholds to manually replace abnormal pixels, risk severe overfitting to specific background levels and inadvertently discarding faint sources near the noise limit \citep{tdr_method}.

Beyond these, recent work has explored conditional denoising diffusion models for radio-astronomical sky estimation and source extraction, showing improvements over classical algorithms \citep{radioDDM, radioDDRM}. Normalizing flow models have also been proposed for denoising density estimations in specific stellar survey data \citep{normalizingflows}. Furthermore, physics-informed neural networks have demonstrated success in removing telescope beam effects \citep{beam_removal}. These studies often focus on specific astronomical targets, rely indirectly on stacked data for evaluation, or rely on specialized data priors which restricts generalizability to broader optical or near-infrared astronomical imaging. 

Other recent advancements include spatiotemporal denoising techniques that leverage heavy transformer architectures across multiple frames \citep{asteris}. This dependency limits their utility since multi-exposure data is frequently unavailable, and the immense parameter count drastically increases computational costs. Similarly, Deep Image Prior (DIP) approaches adapted for multi-frame restoration require solving a new iterative optimization problem for every image sequence, making them computationally prohibitive for large-scale surveys and restricting their application primarily to co-addition rather than isolated single-frame denoising \citep{astroclearnet}.

A comprehensive assessment of how effectively these methods improve the detection of faint, diverse objects (galaxies, stars) in typical survey images, evaluated using metrics like detection completeness and false alarm rates, is still needed. Furthermore, the practical application of SURE requires careful consideration of noise parameter estimation in real observational data \citep{SURE}.

This work, AstroSURE, aims to fill this gap. We adapt and systematically compare the N2N and SURE frameworks for denoising general astronomical images from HST and CFHT, using realistic simulations to complement real data analysis. Our primary goal is to assess the extent to which these unsupervised techniques can enhance the reliable detection of faint sources. 
The contribution of AstroSURE is a detection-focused evaluation of N2N and SURE for astronomical image denoising, using synthetic data together with real HST and CFHT observations to assess practical performance and transfer across observing domains.

Therefore, AstroSURE distinguishes itself by: 
This work is limited to detection-oriented evaluation; we do not claim that the denoised images are suitable for precision photometry, shape measurement, or other PSF-sensitive analyses without dedicated validation.

\section{Data Overview}
\label{sec:data}

\begin{figure}[!htb]
\centering
\subfigure[Full Dynamic Range (Log Scale)]{\includegraphics[width=0.47\linewidth]{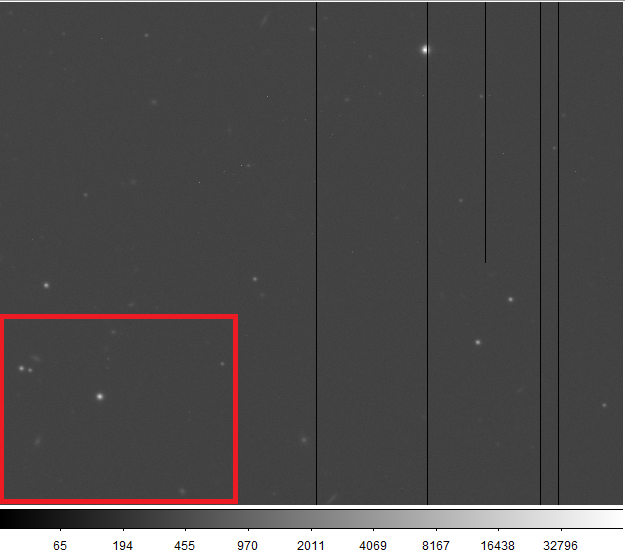}
\label{fig:distribution_log}}
\hfill
\subfigure[Zoom on Faint Structure (Linear Scale)]{\includegraphics[width=0.47\linewidth]{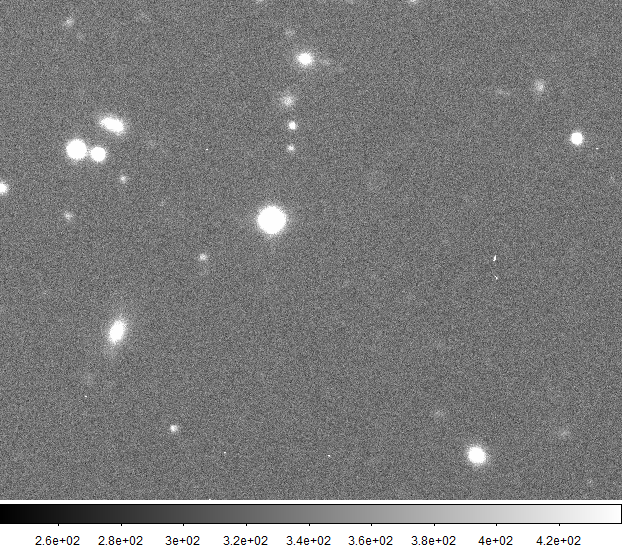}
\label{fig:distribution_zoomed}}
\\
\subfigure[Pixel Value Histogram (Log-Linear)]{\includegraphics[width=1\linewidth]{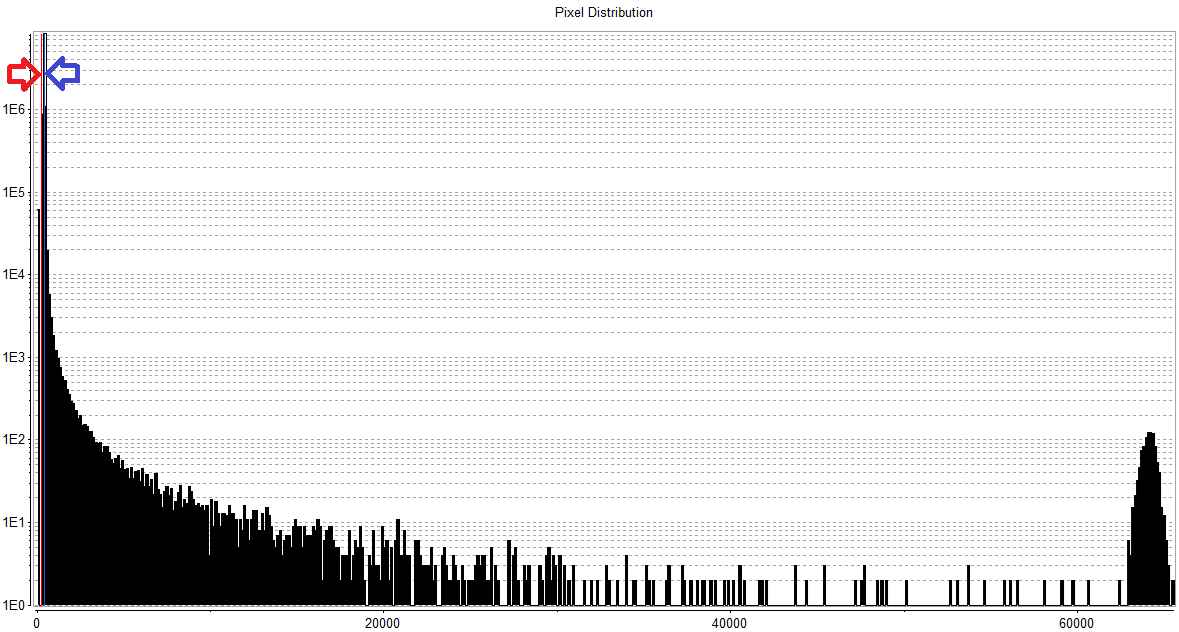}
\label{fig:distribution_histogram}}
\caption{Illustrative characteristics of a MegaPrime exposure. (a) The full dynamic range displayed with a logarithmic stretch, revealing both bright sources and faint nebulosity. (b) A zoomed-in region highlighting faint sources near the background noise level, shown with a linear stretch chosen to emphasize noise properties. (c) Histogram of pixel intensities, showing the dominant background peak and the extended tail from astronomical objects. Arrows indicate the approximate regimes dominated by readout noise and photon statistics.}
\label{figure:distribution}
\end{figure}

Astronomical datasets present unique challenges for image processing algorithms due to their inherent characteristics. Pixel value distributions in astronomical images differ markedly from those in natural-scene imagery. Sensors used in astronomy often exhibit a high dynamic range, capturing a vast range of intensities from faint diffuse structures at the noise limit to bright, potentially saturated, stellar cores. The resulting pixel intensity histograms are heavily skewed, with a dominant peak at the sky background level and a long tail representing astrophysical sources, contrasting with the more diverse set of intensity distributions often found in natural scenes. These properties necessitate careful consideration in the design and application of denoising techniques. For visualization purposes throughout this paper, images are displayed using common astronomical scaling techniques such as percentile-based scaling or Z-scale \citep{astropy:2013, astropy:2018, astropy:2022}. This scaling is applied for display only and does not affect the underlying data used in our analyses. An illustrative example of a raw exposure captured at the CFHT with MegaCam targeting a deep extragalactic field is provided in \autoref{figure:distribution}.

As discussed, to rigorously evaluate and compare unsupervised and supervised denoising techniques, we utilize synthetic data where ground-truth images are actually available. The generation and characteristics of this synthetic dataset are detailed in \autoref{sec:synthetic_data}, and in \autoref{sec:raw_data}, we describe the real observational datasets from HST and CFHT used in our study.

\subsection{Noise Model} \label{sec:synthetic_data}

Contaminants in astronomical imaging can be broadly classified into local and global. Local contaminants, such as cosmic ray hits, satellite trails, diffraction spikes, persistence effects from prior exposures, pixel saturation, and detector defects (e.g., 'hot' or 'dead' pixels, which can manifest as isolated pixels, clusters, or entire rows/columns), affect localized regions of an image. While these can significantly corrupt individual pixels, the majority of the image often remains unaffected. Global contaminants, in contrast, are image-wide imperfections. These include Gaussian-distributed readout noise from detector electronics, Poisson-distributed photon shot noise inherent to the quantum nature of light, and quantization noise from the analog-to-digital conversion process. The low-photon-count signal observations targeted in this work are typically affected by a combination of both local and global contaminants \citep{maximask, astronomy_poisson_denoising}. Our primary focus in this paper is on addressing the challenges posed by unstructured, global contaminants.

We first briefly review the comprehensive noise model implemented in \texttt{GalSim} \citep{galsim}, which underpins our synthetic data generation. A schematic of this model is provided in Appendix~\ref{sec:noise_formation}. The process begins with a noiseless, ideal image based on the input object catalogue. This astrophysical scene is then subjected to simulated photon collection by a CCD. This "photon shooting" step considers the object properties, exposure duration, pointing coordinates, and filter band-pass. Inherent photon shot noise (Poisson distributed) from the sources themselves is incorporated. Additional Poisson noise arises from the sky background, which is a composite of zodiacal light, atmospheric emission (for ground-based), and thermal emission from the telescope and instrument. The sky background level depends on factors like stray light, filter choice, pointing, and observation date, while thermal background is sensitive to instrument temperature and exposure time. Reciprocity failure, where the detector's response deviates from linearity with exposure duration, is modeled. Dark current, arising from thermally generated electrons within the detector, is another significant noise source. It is Poisson-distributed, temperature-dependent, and accumulates with exposure time, occurring even in the absence of incident light.

While the \texttt{GalSim} noise model is highly detailed, a more tractable, simplified noise model is often adopted for developing and testing denoising algorithms. Following \citet{astro_unet}, we can model the observed value at each pixel, $\obsRV$, as a sum of several components:
\begin{equation}
    \obsRV = \photonRV + \readNoiseRV + \darkNoiseRV - \quantNoiseRV,
    \label{eq:noise_model}
\end{equation}
where $\photonRV$ represents the true astrophysical signal (in photo-electrons or ADU, depending on the stage of processing being modeled). The subsequent terms are noise contributions: $\readNoiseRV$ is the readout noise, $\darkNoiseRV$ is the dark current noise, and $\quantNoiseRV$ is the quantization noise.
The photon signal $\photonRV$ (originating from the source and background) is fundamentally Poisson distributed. If $\lambda_{\rm sig}$ is the mean expected signal (e.g., in photo-electrons), then:
\begin{equation}
    \Pr[\photonRV = k] = \frac{\lambda_{\rm sig}^k e^{-\lambda_{\rm sig}}}{k!}, \quad k \geq 0.
    \label{eq:noise_shot}
\end{equation}
Readout noise, $\readNoiseRV$, is typically well-approximated by a zero-mean Gaussian distribution with variance $\sigma^2_{\rm read}$.
Dark current, $\darkNoiseRV$, is also Poisson distributed. However, for sufficiently high dark current rates or long exposures, the Poisson distribution can be approximated by a Gaussian. This approximation is often made for simplicity. For space-based observatories operating at cryogenic temperatures, dark current is often negligible. For ground-based telescopes, while higher, its contribution relative to readout noise can sometimes be small enough that a Gaussian approximation introduces minimal error in the overall noise model. The variance of the dark current noise scales with exposure time, $t$. Thus, $\darkNoiseRV$ can be modeled as a zero-mean Gaussian with variance $\sigma^2_{\rm dark} = \lambda_{\rm dark} t$, where $\lambda_{\rm dark}$ is the mean dark current rate (e.g., electrons/pixel/second), a parameter typically found in instrument handbooks \citep{astro_unet}.
Quantization noise, $\quantNoiseRV$, arises from the digitization process. For high-resolution analog-to-digital converters (e.g., 16-bit), this noise is often modeled as being uniformly distributed over $[-r/2, r/2]$, where $r$ is the width of a quantization step (e.g., 1 ADU). Its impact is often minor compared to other noise sources in low-light astronomical imaging.

Many variations of \eqref{eq:noise_model} exist in the literature. A common and practical simplification, particularly for denoising applications, models the observed pixel value as the sum of a Poisson-distributed signal component (scaled by a gain factor $\kappa$ if working in ADU) and an aggregated Gaussian noise component:
\begin{equation}
    \obsRV \approx \kappa \cdot P(\lambda_{\rm signal}/\kappa) + N(0, \sigma^2_{\rm total}),
    \label{eq:noise_model_final}
\end{equation}
where $P(\lambda_{\rm signal}/\kappa)$ represents the Poisson-distributed photo-electron count scaled by the gain $\kappa$ (ADU/electron), and $N(0, \sigma^2_{\rm total})$ is a Gaussian distribution representing the combined effects of readout noise, Gaussian-approximated dark current, and potentially other minor Gaussian noise sources. The parameters $\sigma_{\rm total}$ and $\kappa$ are critical and are typically derived from instrument characteristics or estimated from calibration data \citep{physics_based_noise_formation,astro_unet, astro_unet_fake}. The variable $\phi$ can represent either photo-electrons or ADU, depending on whether gain has been applied.


In practice, we adopt the simplified mixed Poisson-Gaussian model in \ref{eq:poisson_gaussian_sure_empirical} as a tractable approximation for denoiser training. In the simplified formulation used here, the sky background is treated separately from the source signal. For the synthetic data used in this work, we subtract an estimated background prior to training, as a large near-constant background can dominate the loss and hinder training stability.

\subsection{Telescope Data} \label{sec:raw_data}

To evaluate denoiser performance on real astronomical observations, we utilize datasets from two facilities:


\paragraph{Canada-France-Hawaii Telescope (CFHT)} We used twenty MegaCam exposures obtained with the r.MP9601 filter as part of the Next Generation Virgo Survey (NGVS)\citep{NGVS1, NGVS2, NGVS3, NGVS4, NGVS5, NGVS6}. The images were processed with the standard CFHT Elixir pipeline, including instrumental-signature removal such as flat-fielding and overscan correction. Each multi-extension FITS file contains 36 CCD frames of size $2112 \times 4644$ pixels. From these, $256 \times 256$ pixel patches were extracted for training and evaluation. Prior to patch extraction, each CCD frame was trimmed to remove detector-edge regions and unusable pixels. Obvious bad pixels were interpolated using values from neighboring pixels. The WCS solution for each frame was refined using CASUTools \citep{casu}, cross-matching sources against the Gaia Data Release 2 (DR2) catalogue \citep{gaia1, gaia2, gaia3}.

\paragraph{Hubble Space Telescope (HST)} A comparable dataset was sourced from the HST Wide Field Camera 3 (WFC3), specifically from observations of the Hubble Deep Fields (e.g., GOODS, HUDF). These data, taken with various filters (primarily F606W and F775W), represent some of the deepest optical/near-IR views. For WFC3/UVIS, each exposure typically consists of a $2 \times 1$ mosaic of CCDs, yielding frames of approximately $4102 \times 4096$ pixels. These data were processed using standard HST calibration pipelines (e.g., CALWF3) by the Space Telescope Science Institute (STScI), including corrections for bias, dark current, flat-fielding, and CTE effects \citep{hst_processing}. Similar to the CFHT data, $256 \times 256$ pixel patches were extracted for our experiments after any necessary pre-processing like cosmic ray rejection.

\section{Methodology}
\label{sec:method}

This section defines the training objectives, network architectures, and optimization procedure used in this work. We then summarize the AstroSURE workflow adopted in the experiments and the evaluation metrics used in the next section.

\subsection{Problem Formulation}
\label{sec:problem}

We evaluate denoisers using several loss functions, tailored to different data availability scenarios. Let $u_i \in \mathbb{R}^n$ represent a clean target image (ground truth) and $y_i \in \mathbb{R}^n$ be a noisy observation thereof, where $n$ is the number of pixels. A denoiser, parameterized by $\theta$, is a function $f_{\theta} : \mathbb{R}^n \rightarrow \mathbb{R}^n$ that maps noisy observations to estimates of the clean target.

\paragraph{Supervised Learning}
In the ideal supervised setting, a dataset of $N$ pairs $\{(u_i, y_i)\}_{i=1}^N$ is available. The model parameters $\theta$ are optimized by minimizing a loss function that measures the discrepancy between the denoised output $f_{\theta}(y_i)$ and the clean target $u_i$. Standard choices include the Mean Squared Error (MSE) and Mean Absolute Error (MAE):
\begin{equation}
    L_{\textrm{MSE}}(\theta) = \frac{1}{N} \sum_{i=1}^{N} \| u_i - f_{\theta}(y_i)\|_{2}^2, \hspace{2em} {\rm and} \hspace{2em}
    L_{\textrm{MAE}}(\theta) = \frac{1}{N} \sum_{i=1}^{N} \| u_i - f_{\theta}(y_i)\|_1.
    \label{eq:supervised_msemae}
\end{equation}
The $L_2$ norm (MSE) penalizes larger errors more heavily, while the $L_1$ norm (MAE) is more robust to outliers.

\paragraph{Learning with Noisy Targets: Noise2Noise}
In astronomical observations, clean target images $u_i$ are generally unavailable. The Noise2Noise (N2N) framework \citep{n2n} provides a strategy for learning with paired noisy observations, $(y_i', y_i)$, of the same underlying clean signal $u_i$, without requiring clean targets. It is based on the premise that if the noise is additive, zero-mean, and independent in $y_i'$ and $y_i$, training a denoiser $f_{\theta}(y_i)$ to predict the other noisy instance $y_i'$ using an MSE loss is equivalent, in expectation, to training with the clean target $u_i$:
\begin{equation}
    L_{\textrm{N2N}}(\theta) = \frac{1}{N} \sum_{i=1}^{N} \|y_i' - f_{\theta}(y_i)\|_{2}^2.
    \label{eq:n2n_mse}
\end{equation}
The core insight is that for deviation-minimizing estimators (like MSE), the learning task $E[y_i' | y_i]$ converges to $E[u_i | y_i]$ if $E[\text{noise}] = 0$. N2N can be effective even with some forms of local, structured noise, provided the zero-mean and independence conditions on the noise instances hold. A practical limitation is the need for aligned pairs of noisy images, which can be challenging to acquire in astronomy.

\paragraph{Self-Supervised Learning: Blind-Spot Networks}
To overcome the paired-image requirement of N2N, several single-image self-supervised methods have been developed, including Noise2Void \citep{noise2void}, Noise2Self \citep{noise2self}, and Noise2Same \citep{noise2same}. These often employ "blind-spot" networks. Such networks predict the value of a target pixel $y_{i,p}$ using only its contextual pixels, explicitly excluding $y_{i,p}$ itself from the receptive field for that prediction. This prevents the network from trivially learning an identity mapping and forces it to learn from the underlying signal structure. The loss is typically an MSE between the masked pixel's actual noisy intensity value and its prediction:
\begin{equation}
    L_{\textrm{self-sup}}(\theta) = \frac{1}{N} \sum_{i=1}^{N} \sum_{p \in \mathcal{M}_i} \|y_{i,p} - (f_{\theta}(y_{i,masked}
    ))_{p}\|_{2}^2,
    \label{eq:self_supervised_loss}
\end{equation}
where $\mathcal{M}_i$ is the set of masked pixels in image $y_i$, and $y_{i,masked}$ indicates that $f_\theta$ predicts pixels in $\mathcal{M}_i$ without directly seeing them.

\paragraph{Unsupervised Learning: Stein's Unbiased Risk Estimator (SURE)}
SURE provides an analytical estimate of the mean-squared risk of a denoiser $f_{\theta}(y)$ without access to the clean signal $u$, under suitable assumptions on the noise model and the denoiser. For an observation $y \in \mathbb{R}^m$ (where $m$ is the number of pixels) corrupted by additive zero-mean Gaussian noise with known variance $\sigma^2$ (i.e., $y = u + \epsilon$, $\epsilon \sim \mathcal{N}(0, \sigma^2 I_m)$), the SURE loss for a single image is:
\begin{equation}
    L_{\textrm{SURE}}(f_{\theta}(y)) = \|y - f_{\theta}(y)\|_{2}^2 - m\sigma^2 + 2\sigma^2 \mathrm{div}_{y} f_{\theta}(y).
    \label{eq:sure_gaussian_exact}
\end{equation}
The term $\mathrm{div}_{y} f_{\theta}(y) = \sum_{k=1}^m \frac{\partial (f_{\theta}(y))_k}{\partial y_k}$ is the divergence of the denoiser $f_{\theta}$ with respect to its input $y$. The first term measures the discrepancy between the observation and its denoised version, while the divergence term corrects for bias. The overall loss for a dataset is $L_{\textrm{SURE}}(\theta) = \frac{1}{N} \sum_{i=1}^{N} L_{\textrm{SURE}}(f_{\theta}(y_i))$.

Calculating the divergence analytically is often intractable for complex denoisers like neural networks. A Monte Carlo approximation is commonly used \citep{monte_carlo}:
\begin{equation}
    \mathrm{div}_{y} f_{\theta}(y) \approx \mathbb{E}_{b \sim \mathcal{N}(0,I_m)} \left[ b^T \frac{f_{\theta}(y+\tau b) - f_{\theta}(y)}{\tau} \right],
    \label{eq:divergence_mc_general}
\end{equation}
for a small $\tau > 0$. In practice, this expectation is often approximated using a single random perturbation vector $b$:
\begin{equation}
    \mathrm{div}_{y_i} f_{\theta}(y_i) \approx \frac{1}{\tau}b_i^{T}(f_{\theta}(y_i+\tau b_i) - f_{\theta}(y_i)),
    \label{eq:divergence_approx}
\end{equation}
where $b_i \sim \mathcal{N}(0,I_m)$. Substituting this into \eqref{eq:sure_gaussian_exact} yields the empirical SURE loss for Gaussian noise for a single image $y_i$ (denoted $L_{\textrm{SURE-G}}$):
\begin{equation}
    L_{\textrm{SURE-G}}(\theta; y_i) = \|y_i - f_{\theta}(y_i)\|_{2}^2 - m\sigma^2 + \frac{2 \sigma^2}{\tau} b_i^T(f_{\theta}(y_i + \tau b_i) - f_{\theta}(y_i)).
    \label{eq:gaussian_sure_empirical}
\end{equation}
The total loss is averaged over the dataset: $L_{\textrm{SURE-G}}(\theta) = \frac{1}{N}\sum_{i=1}^{N} L_{\textrm{SURE-G}}(\theta; y_i)$.

SURE has been extended to other noise models, including Poisson and mixed Poisson-Gaussian noise \citep{esure, general_sure, poisson_gaussian_sure, sure_mri, sure_extension, sure_deep, sure_deep2}.
For Poisson noise (denoted $L_{\textrm{SURE-P}}$) with mean (and variance) parameter $\lambda_p$ (assuming $y_i \sim \text{Poisson}(\lambda_p u_i)$ or similar, where $\lambda_p$ might be a global scaling or pixel-wise rate), an approximate SURE variant using a similar Monte Carlo divergence approximation is \citep{SURE, esure, general_sure, ei}:
\begin{equation}
\begin{split}
    L_{\textrm{SURE-P}}(\theta; y_i) = \|y_i - f_{\theta}(y_i)\|_{2}^2 - \mathbf{1}^T y_i + \frac{2}{\tau} {(b_i \odot y_i)}^T (f_{\theta}(y_i + \tau b_i) - f_{\theta}(y_i)).
    \label{eq:poisson_sure_empirical}
\end{split}
\end{equation}
Here, $b_i$ is a vector whose elements are drawn from a Bernoulli distribution (e.g., $\pm 1$ with equal probability), $\mathbf{1}^T y_i$ is the sum of pixel values in $y_i$, and $\odot$ denotes element-wise multiplication. The parameter $\lambda_p$ is implicitly handled or absorbed into $y_i$ or $f_\theta$ scaling in some formulations; here it appears to be assumed that the mean of the noise is related to $y_i$ itself for the divergence term.

For mixed Poisson-Gaussian noise (denoted $L_{\textrm{SURE-PG}}$), where observations are subject to both Poisson photon statistics and additive Gaussian readout noise (variance $\sigma_g^2$), a more complex SURE formulation is required. Following \citet{rei}, an approximate loss is:
\begin{equation}
\begin{split}
    L_{\textrm{SURE-PG}}(\theta; y_i) = & \|y_i - f_{\theta}(y_i)\|_{2}^2 - \mathbf{1}^T y_i - m\sigma_g^2 \\
    & + \frac{2}{{\tau}_1} {(b_i \odot (y_i + \sigma_g^2 \mathbf{1}))}^T (f_{\theta}(y_i + {\tau}_1 b_i) - f_{\theta}(y_i)) \\
    & - \frac{2 \sigma_g^2}{{\tau}_2} c_i^T \left( f_{\theta}(y_i + {\tau}_2 c_i) + f_{\theta}(y_i - {\tau}_2 c_i) - 2f_{\theta} (y_i)\right).
    \label{eq:poisson_gaussian_sure_empirical}
\end{split}
\end{equation}
Here, $y_i$ is assumed to be scaled such that its Poisson component has a rate related to its own values (a common approximation). $b_i$ is a zero-mean isotropic Gaussian vector, $c_i$ is a Bernoulli-distributed vector ($\pm 1$), $\mathbf{1}$ is a vector of ones, and ${\tau}_1, {\tau}_2$ are small positive step sizes for the Monte Carlo approximations of the first and second-order derivative terms related to divergence. The term $(y_i + \sigma_g^2 \mathbf{1})$ in the Poisson-related divergence part aims to approximate the variance of the underlying Poisson-Gaussian signal.
The specific empirical SURE loss ($L_{\textrm{SURE-G}}$, $L_{\textrm{SURE-P}}$, or $L_{\textrm{SURE-PG}}$) used will be clear from the context of the noise model.

\subsection{Network Architectures}\label{sec:architectures}
We evaluated several standard denoising architectures and retained a modified U-Net as the primary model for the experiments that follow.

\paragraph{DnCNN}
The Denoising Convolutional Neural Network (DnCNN) \citep{dncnn}, typically trained in a supervised manner, is a Fully Convolutional Network (FCN). FCNs exclusively use convolutional layers, enabling them to process inputs of arbitrary size and produce outputs of the same spatial dimensions. DnCNN typically employs $3 \times 3$ convolutional filters, batch normalization, and ReLU activation functions. It is trained to predict the residual noise, which is then subtracted from the noisy input to yield the denoised image. Despite its relative simplicity and age, DnCNN remains a widely used baseline.

\paragraph{U-Net}
The U-Net architecture \citep{unet}, originally developed for biomedical image segmentation, is also highly effective for denoising. Its characteristic U-shape consists of an encoder path that progressively downsamples the input to capture contextual information, and a decoder path that symmetrically upsamples to reconstruct the output. Skip connections link corresponding encoder and decoder layers, facilitating the flow of high-resolution feature information from the encoder to the decoder; this allows the network to learn multi-scale features important for distinguishing noise from signal while preserving fine details.
We consider a "vanilla" U-Net similar to that in \citet{astro_unet}. For a $256 \times 256$ input, the encoder typically uses blocks of two convolutions (with Leaky ReLU activations) followed by max-pooling, reducing spatial dimensions while increasing feature channels. The decoder uses transposed convolutions for upsampling, followed by convolution blocks and concatenation with features from skip connections. A final $1 \times 1$ convolution maps features to the output image.

\paragraph{Modified U-Net}
We use a modified U-Net architecture inspired by \citet{n2n} and similar to models previously found effective for astronomical image denoising \citep{unet_improved}. Relative to a vanilla U-Net, the main changes are in the channel allocation across encoder and decoder stages. This architecture is shown in \autoref{figure:unet} and is used as the primary model in the remainder of the paper

\paragraph{Restormer}
The Restoration Transformer (Restormer) \citep{restormer} adapts the Transformer architecture, successful in natural language processing, for image restoration tasks. It aims to capture long-range dependencies efficiently in 2D data. Restormer employs mechanisms like Multi-Dconv Head Transposed Attention (MDTA) and Gated-Dconv Feed-Forward Networks (GDFN) within its transformer blocks to model both local and global contexts effectively. While achieving state-of-the-art performance on many benchmarks, Restormer typically has a significantly larger number of parameters (e.g., ~25 million) compared to U-Net variants (~1 million for our modified U-Net), leading to increased computational cost for training and inference.

\subsection{Training Details} \label{sec:training}
For robust and comparable training across models, we primarily use the Adam optimizer \citep{adam_optimizer} with parameters $\beta_1=0.9$, $\beta_2=0.99$, and $\epsilon=10^{-8}$. The AMSGrad extension \citep{amsgrad} is incorporated to potentially improve convergence.
The only exception is for the Restormer model, which is trained following author recommendations: Adam with weight decay $10^{-4}$, $\beta_1=0.9$, $\beta_2=0.999$, $\epsilon=10^{-8}$, without AMSGrad \citep{restormer}.

The initial learning rate is set to $0.002$. A learning rate scheduler reduces the rate by a factor of two if the validation loss does not improve for $20$ consecutive epochs. Models are typically trained for up to $200$ epochs, while Restormer is trained for $50$ epochs as suggested by the original paper. Early stopping is employed, i.e., training terminates if the validation loss does not improve by at least $0.1\%$ for $10$ epochs after the last learning rate reduction. Each epoch involves processing the entire training dataset in mini-batches.

\subsection{AstroSURE Pipeline} \label{sec:pipeline}

\begin{figure}[!htb]
\centering
\includegraphics[width=1\linewidth]{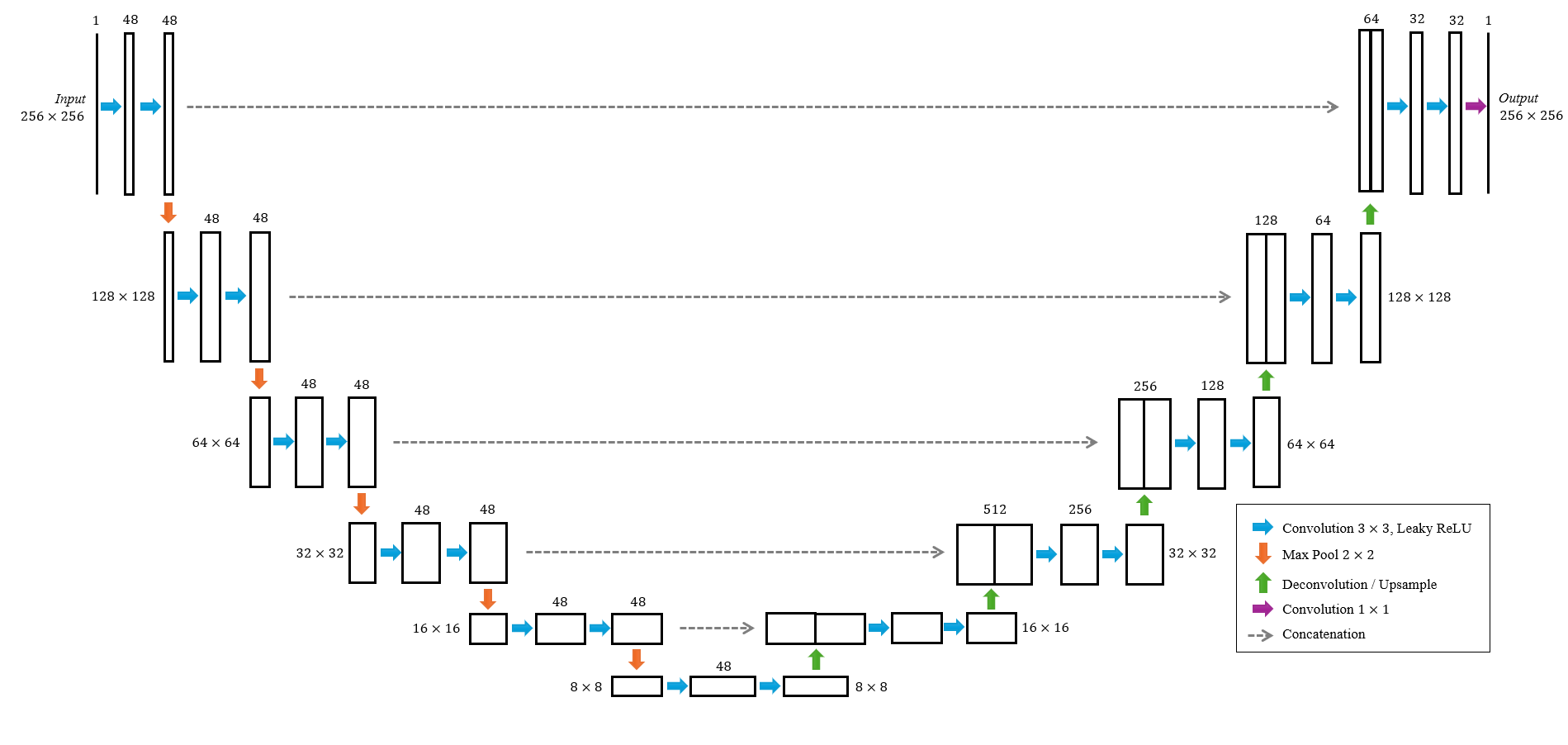}
\caption{Architecture of the modified U-Net}
\label{figure:unet}
\end{figure}

We use the term AstroSURE to refer to the practical workflow adopted in this paper for training denoisers on astronomical images without clean targets. In the real-data experiments, the primary configuration uses the modified U-Net architecture (\autoref{figure:unet}) together with the mixed Poisson-Gaussian SURE loss in \eqref{eq:poisson_gaussian_sure_empirical}. When available, the network may be initialized from a model trained on a related astronomical domain, for example through Noise2Noise training on paired noisy data or through simulations.

The workflow in this study is as follow:
\begin{enumerate}
    \item \textbf{Data Preparation:} Full-frame astronomical images are divided into non-overlapping $256 \times 256$ pixel patches. 
    Preprocessing and scaling are applied as specified for each experiment.
    \item \textbf{Training:} The network is trained patch-wise using the selected loss function. For SURE-based training, the required noise parameters are estimated from instrument characteristics or from the data.
    \item \textbf{Validation:} Validation is performed on held-out data using the metrics appropriate to each setting. When relevant, denoised outputs are transformed back to the original data scale before evaluation
    \item \textbf{Inference:} A new image is patched, denoised patch-by-patch, and reconstructed into a full-frame output for downstream analysis
\end{enumerate}
In this work, AstroSURE is evaluated primarily in a detection-oriented setting rather than as a general-purpose image-restoration framework.

\subsection{Evaluation Metrics}
\label{sec:metrics}
We employ several metrics to evaluate denoiser performance. For a true clean image $u$ and its denoised estimate $\tilde{u}$, both of size $m \times n$ pixels:

\begin{itemize}
    \item \textbf{Mean Squared Error (MSE):} $\textrm{MSE} = \frac{1}{mn} \sum_{i=1}^{m} \sum_{j=1}^{n} (\tilde{u}_{ij} - u_{ij})^2$.
    \item \textbf{Mean Absolute Error (MAE):} $\textrm{MAE} = \frac{1}{mn} \sum_{i=1}^{m} \sum_{j=1}^{n} |\tilde{u}_{ij} - u_{ij}|$.
    \item \textbf{Signal-to-Noise Ratio (SNR):} Generally defined as $\textrm{SNR} = P_{\textrm{Signal}} / P_{\textrm{Noise}}$. In image processing, this is often computed based on signal variance to noise variance, or mean signal level to noise standard deviation in a region. 
    \item \textbf{Peak Signal-to-Noise Ratio (PSNR):} Common in image processing, defined as:
    \begin{equation}
        \textrm{PSNR}_{\textrm{dB}} = 10 \log_{10} \left(\frac{\textrm{MAX}^2}{\textrm{MSE}} \right) = 20 \log_{10}(\textrm{MAX}) - 10 \log_{10}(\textrm{MSE}).
        \label{eq:psnr}
    \end{equation}
    MAX is the maximum possible pixel value. For typical 16-bit astronomical data, MAX = $2^{16}-1 = 65535$. Thus, $\textrm{PSNR}_{\textrm{dB}} \approx 20 \log_{10}(65535) - 10 \log_{10}(\textrm{MSE}) \approx 96.33 - 10 \log_{10}(\textrm{MSE})$.

    \item \textbf{Kullback-Leibler (KL) Divergence:} To compare the pixel intensity distributions, we treat normalized images as probability distributions. Given two non-negative images $\tilde{u}$ (test) and $u$ (reference), let $P_{ij} = \tilde{u}_{ij} / \sum_{k,l} \tilde{u}_{kl}$ and $Q_{ij} = u_{ij} / \sum_{k,l} u_{kl}$. The KL divergence of $P$ from $Q$ is:
    \begin{equation}
        D_{KL} (P||Q) = \sum_{i=1}^{m} \sum_{j=1}^{n} P_{ij} \log \left(\frac{P_{ij}}{Q_{ij}}\right).
        \label{eq:kl_standard}
    \end{equation}
    This can be rewritten in terms of image pixel values $\tilde{u}_{ij}$, $u_{ij}$ and their sums $\tilde{Z} = \sum_{k,l} \tilde{u}_{kl}$, $Z = \sum_{k,l} u_{kl}$:
    \begin{equation}
        D_{KL} (\tilde{u}||u) = \sum_{i=1}^{m} \sum_{j=1}^{n} \frac{\tilde{u}_{ij}}{\tilde{Z}} \log \left(\frac{\tilde{u}_{ij}/ \tilde{Z}}{u_{ij}/Z} \right) = \sum_{i=1}^{m} \sum_{j=1}^{n} \frac{\tilde{u}_{ij}}{\tilde{Z}} \left( \log \left(\frac{\tilde{u}_{ij}}{u_{ij}}\right) + \log \left(\frac{Z}{\tilde{Z}}\right) \right).
        \label{eq:kl_image}
    \end{equation}
    $D_{KL} \ge 0$, with $D_{KL}=0$ if and only if $P=Q$. This metric helps assess if denoising preserves the overall statistical properties of the image.

    \item \textbf{Correct Detection Rate (CDR):} Ratio of correctly detected true objects to the total number of true objects.
    \item \textbf{False Alarm Rate (FAR):} Ratio of falsely detected objects to the total number of detections made.

    \item \textbf{Unsupervised PSNR (uPSNR):} Calculated using an unsupervised MSE (uMSE) estimate \citep{unsupervised_metrics}. Given a noisy observation $y$ and its denoised version $f(y)$, uMSE requires three additional independent noisy instances ($y', y'', y'''$) of the same underlying scene:
    \begin{equation}
        \textrm{uMSE} = \frac{1}{mn} \sum_{i,j} \left( (y'_{ij} - f(y_{ij}))^2 - \frac{(y''_{ij} - y'''_{ij})^2}{2} \right).
        \label{eq:umse}
    \end{equation}
    The subtraction aims to cancel noise terms, providing an estimate of the true MSE. For single noisy images, $y', y'', y'''$ can be approximated by grid-sampling sub-images, though this assumes local pixel similarity, which may introduce bias in heterogeneous astronomical fields.
    \item \textbf{Naturalness Image Quality Evaluator (NIQE):} \citep{niqe} A no-reference metric that measures deviations from statistical regularities observed in a corpus of natural, pristine images. Lower NIQE scores indicate better perceptual quality.
\end{itemize}

For the CDR and FAR, detections are performed on full-field images. Then the detected catalogues are cross-matched with reference catalogues (ground truth for synthetic data, or curated catalogues from NGVS for the real data experiments) using STILTS \citep{stilts}, with a matching radius (e.g., 1 arcsecond). The uPSNR and NIQE metrics are also specifically used in scenarios where no ground truth or reference was available to evaluate or monitor the training process.

Visual evaluation of denoised images, residuals ($y - f_{\theta}(y)$), and error maps ($u - f_{\theta}(y)$ where available) also forms an important part of our qualitative assessment.

\section{Experiments} \label{sec:experiments}
In this section, we evaluate the proposed denoising framework on synthetic and real astronomical images. We first compare several network architectures on synthetic data in order to select a practical model for the remaining experiments. We then compare training schemes under different data-availability settings and assess their impact on image quality and, most importantly, source detection performance.

\subsection{Comparison of Network Architectures} \label{sec:architecture_experiment}

We first compare several architectures in a supervised setting on synthetic data in order to select a practical model for the subsequent experiments. This setting provides an upper-bound reference, since both noisy inputs and clean targets are available. \autoref{tab:architectures} shows that no single architecture dominates across all metrics. Restormer achieves the strongest performance on several signal-based metrics, but at substantially higher computational cost, while the modified U-Net remains competitive and is considerably more efficient to train. Given this trade-off, we adopt the modified U-Net as the primary architecture in the experiments that follow.

For all the architectures that allow training for $L1$ and $L2$ losses, the $L1$ training loss mostly outperforms the $L2$ loss. We discuss this more deeply in the \autoref{sec:schemes}.

\begin{table}[!htb]
\caption{Performance comparison of deep neural network architectures trained with supervised settings. Lower numbers are preferred in all columns except in the PSNR column. The DnCNN-17-64 and DnCNN-20-128 represent the DnCNN networks with 17 and 20 stacked layers and network input sizes of 64 and 128, respectively.}
\label{tab:architectures}
\small
\renewcommand{\arraystretch}{1.3}
\begin{tabular}{blcbccc|c|cb}
\bighline
\textbf{Architecture} & \textbf{Loss} & \textbf{MAE} & \textbf{MSE} & \textbf{PSNR (dB)} & \textbf{KL Div.} & \textbf{Training Time} \\
\hline
\multirow{2}{*}{\textbf{Vanilla U-Net}}         & L1 & 1.04 & 247.95 & 81.52 & 2.18 & 6h10m \\
                                               & L2 & 1.81 & 37.85 & 84.49 & 3.35 & 6h5m \\
\hline
\multirow{2}{*}{\textbf{U-Net-TransConv}} & L1 & 0.47 & 42.52 & 88.78 & \textbf{0.80} & 4h \\
                                               & L2 & 0.94 & 18.42 & 87.43 & 3.46 & 4h \\
\hline
\multirow{2}{*}{\textbf{U-Net-Upsample}}        & L1 & 0.51 & 35.44 & 88.25 & 0.84 & 3h50m \\
                                               & L2 & 3.52 & 26.96 & 86.04 & 3.22 & \textbf{3h48m} \\
\hline
\multirow{2}{*}{\textbf{DnCNN-17-64}}          & L1 & 1.32 & 20.24 & 91.45 & 2.97 & 6h51m \\
                                               & L2 & 1.68 & 21.33 & 89.83 & 11.07 & 6h40m \\
\hline
\multirow{2}{*}{\textbf{DnCNN-20-128}}         & L1 & 1.57 & 32.11 & 84.60 & 6.35 & 6h50m \\
                                               & L2 & 2.11 & 28.49 & 85.84 & 6.49 & 7h1m \\
\hline
\textbf{Restormer}                             & L1 & \textbf{0.46} & \textbf{10.80} & \textbf{93.36} & 3.32 & 83h23m \\
\hline\hline
\textbf{Noisy Image}                           &  & 21.26 & 836.91 & 68.39 & 16.40 & - \\
\bighline
\end{tabular}
\end{table}

\begin{figure}[!htb]
\centering
\includegraphics[width=0.85\linewidth]{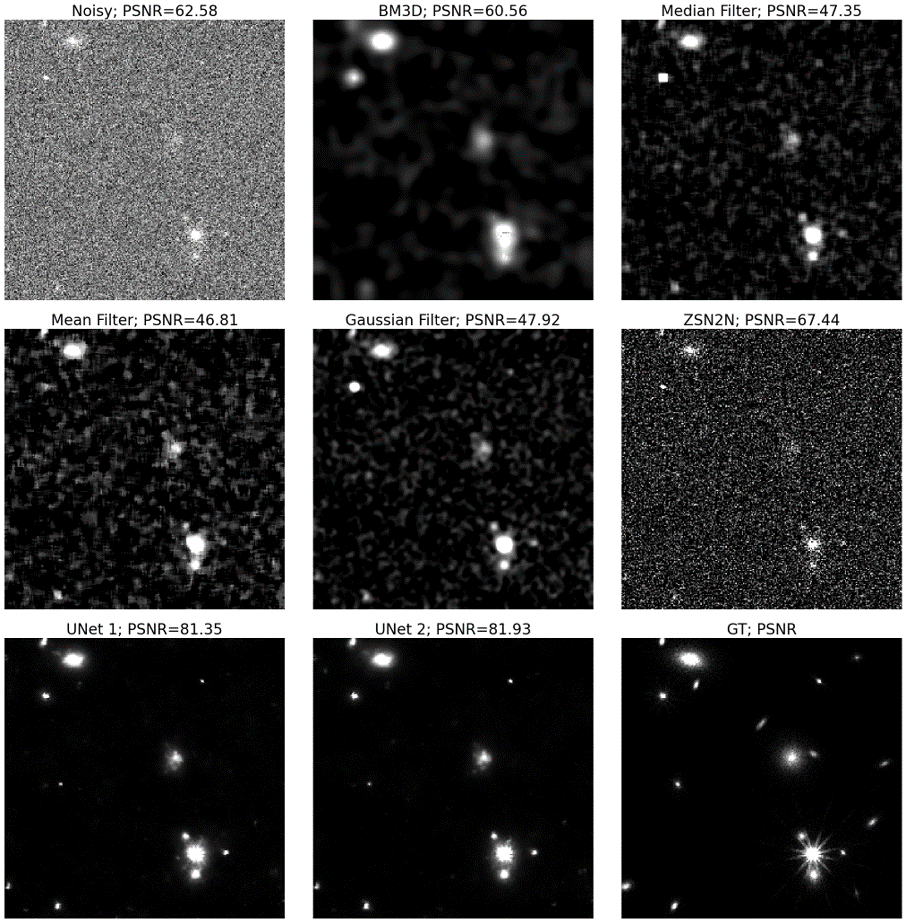}
\caption{Reconstruction result and PSNR of a noisy simulated image; U-Net 1 and 2 are modified U-Nets with Upsampling layers trained using Noise2Clean and Noise2Noise training schemes, respectively.}
\label{figure:result_classic_vs_deep}
\end{figure}

\autoref{figure:result_classic_vs_deep} visualizes the denoised outputs of a synthetic noisy input contaminated with Poisson and Gaussian noise denoised by  U-Net models trained with Noise2Clean and Noise2Noise training strategies, compared to alternative denoising approaches, including several classic filters, Block Matching and 3D Filtering approach, and   cc Zero-shot Noise2Noise. The title of each frame shows the name of the denoising algorithm/architecture and the PSNR of the denoised image. It could be seen that even the best-performing classic approach of BM3D is not able to eliminate the noise and keep the sources intact. The other dataset-free approach of ZSN2N \cite{zsn2n} can also not fully remove the noise components, as seen by the PSNR values and visual quality of images.

As discussed in \autoref{sec:background}, dataset-free approaches typically rely on finding and aggregating similar patches within a single image to suppress noise. While this strategy can be effective for natural images with repeating structures, it falls short for astronomical images, which often feature only a few isolated sources against a mostly dark background. Due to the unique morphology and intensity distribution of each galaxy, these images rarely exhibit internal redundancy which such methods depend on. This limitation highlights the inadequacy of relying solely on information from a single noisy frame. Therefore, it becomes essential to adopt denoising methods that leverage additional context or statistical priors learned from datasets.

\subsection{Effectiveness of Training Schemes} \label{sec:schemes}

In this set of experiments, we compare denoisers trained without clean targets, using the selected architecture and pre-processing pipeline. These experiments are intended to assess how well different weakly supervised or unsupervised training schemes perform under realistic data-availability constraints.

\subsubsection{Simplified Noise model}

\begin{figure}[!htb]
\centering
\includegraphics[width=0.7\linewidth]{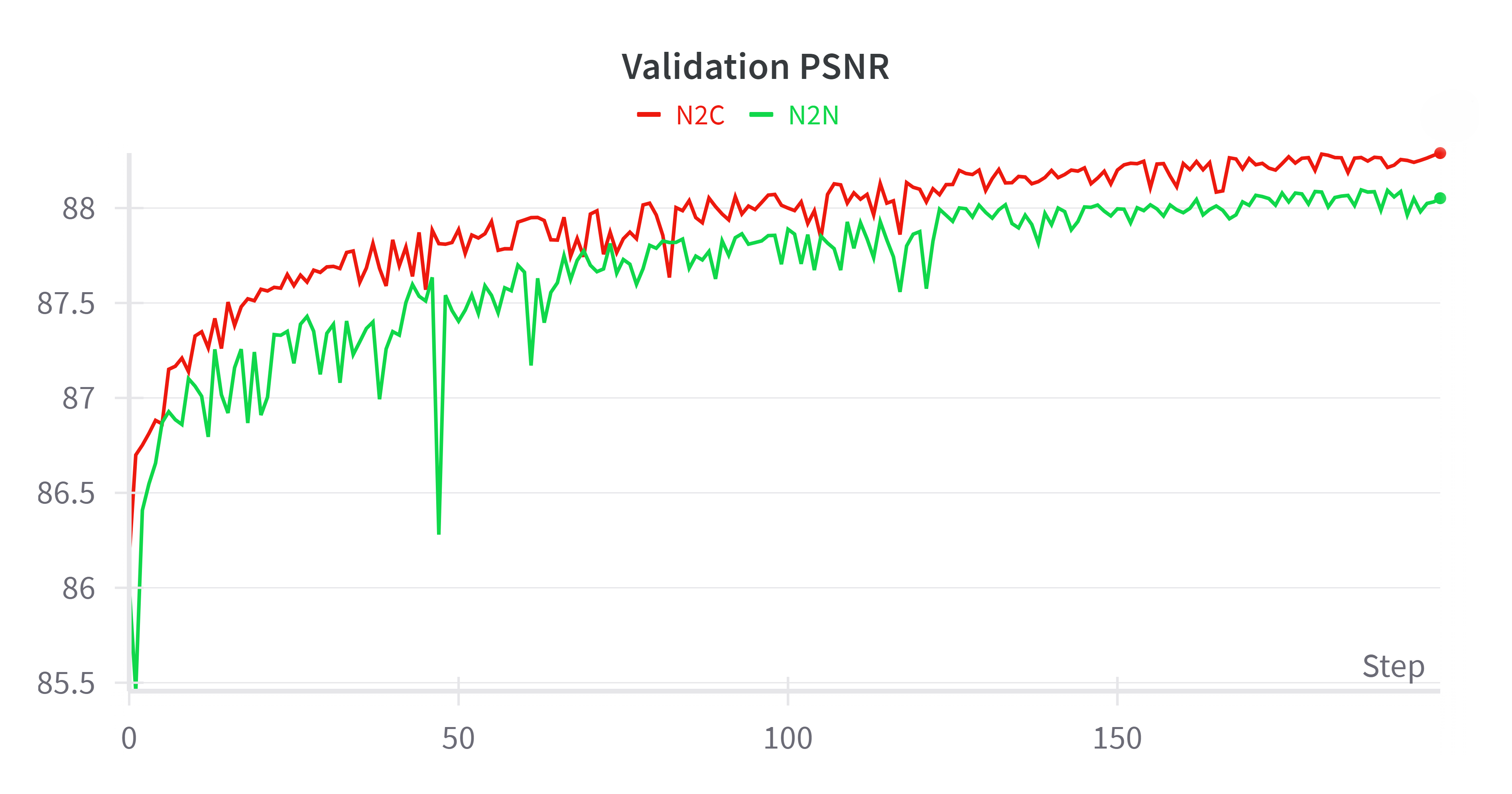}
\caption{Validation PSNR during training of the modified U-Net trained with Noise2Clean and Noise2Noise settings; The Y-axis of the plot is in dB.}
\label{figure:training}
\end{figure}

We first compare training with clean targets, paired noisy targets, and single noisy images using the modified U-Net. \autoref{figure:training} shows that Noise2Noise training closely tracks supervised Noise2Clean training in validation PSNR, despite not requiring clean targets. We then extend the comparison to methods that operate on a single noisy image.

\autoref{tab:simple_noise} illustrates the in-depth comparison of inference on the test set between different denoisers. Each row shows the results of the modified U-Net with upsampling layers trained with the specified training algorithm and the loss that was minimized. The bold numbers accentuate the best result achieved for each metric. 

The results show that Noise2Noise performs comparably to Noise2Clean across the main metrics, while SURE-based denoising remains competitive in the setting where only single noisy images are available.
This can be attributed to the implicit regularization effect of learning from independently corrupted pairs, which helps reduce overfitting and better captures the true noise distribution, especially when “clean” labels are imperfect. Previous work across various imaging domains has noted this phenomenon, underscoring the robustness and generalization capabilities of noise-to-noise supervision. SURE-based denoising shows nearly comparable performance, proving a suitable solution in settings where the data for Noise2Clean and Noise2Noise is unavailable.

\begin{table}[!htb]
\centering
\caption{Performance comparison of denoisers trained with simulated data and simple noise model. The network used in all approaches is the same modified U-Net architecture with Upsampling layers.}
\label{tab:simple_noise}
\footnotesize
\renewcommand{\arraystretch}{1.5}
\begin{tabular}{blbcccc|cc|cccb}
\bighline
Denoiser & 
MAE & 
MSE &
PSNR (dB)& 
SNR (dB) & 
KL Div. & 
NIQE & 
Detection Count & 
FAR (\%) &
CDR (\%) \\
\bighline
N2C L1     &\textbf{0.78} & \textbf{25.37} & \textbf{82.70} & \textbf{32.56} & 0.09 & 20.82 & 2140.61 & 0.49 & \textbf{56.90} \\
N2C L2     &0.99 & 29.66 & 82.03 & 31.89 & 0.12 & 17.54 & 2129.06 & 0.50 & 56.5 \\
N2N L1     &0.90 & 27.92 & 82.29 & 32.14 & \textbf{0.08} & 21.24 & 2122.98 & 0.30 & 56.48 \\
N2N L2     &0.99 & 30.92 & 81.84 & 31.70 & 0.12 & 17.93 & 2123.82 & 0.54 & 56.38 \\
N2Se L1    &1.68 & 7436.37 & 57.88 & 5.65 & 0.30 & 20.90 & 2006.20 & \textbf{0.13} & 53.32 \\
N2Se L2    &3.52 & 8657.51 & 57.24 & 5.39 & 1.26 & 26.77 & 2101.41 & 2.43 & 54.44 \\
N2Sa L2    &24.62 & 1114.13 & 66.90 & 16.95 & 16.08 & 24.11 & - & - & - \\
SURE L2    &1.20 & 32.47 & 81.63 & 31.47 & 0.20 & \textbf{13.38} & 1964 & 0.32 & 52.31 \\
\hline \hline
Noisy Image & 24.61 & 1112.45 & 66.98 & 16.96 & 16.40 & 24.18 & 2044.24 & 3.32 & 50.41 \\
\bighline
\end{tabular}
\end{table}

The Noise2Same training strategy failed to produce a functional model in our experiments. In fact, it negatively impacted the background, leading to failure in the subsequent detection algorithm. We experimented with various learning rates, masking strategies, and invariance scalars, but none led to successful training. We believe this is due to the small size of the objects and the near-zero intensity of most pixels in our images. Since the Noise2Same loss is computed only over a subset of masked pixels, the resulting gradients are too weak or uninformative to guide the model effectively. As noted in \citet{noise2same}, this method relies on the presence of sufficiently large and strong signals within the image, which our dataset lacks.

The original Noise2Self method using the $L_2$ loss does not perform well in our setting. While it marginally improves detection, it significantly underperforms on signal-related metrics—often suppressing both background noise and the signal itself. This likely stems from the limited supervision provided by computing loss on only a small subset of pixels, combined with the unique characteristics of astronomical data, such as sharp intensity variations between neighboring pixels. Across our experiments, models trained with $L_1$ loss consistently outperform their $L_2$ counterparts. The $L_1$ loss is more robust to outliers and better preserves bright astronomical features, minimizing noise without compromising signal integrity. It also captures subtle intensity differences more effectively, resulting in higher-quality denoising.

\begin{figure}[!htb]
\centering
\includegraphics[width=0.85\linewidth]{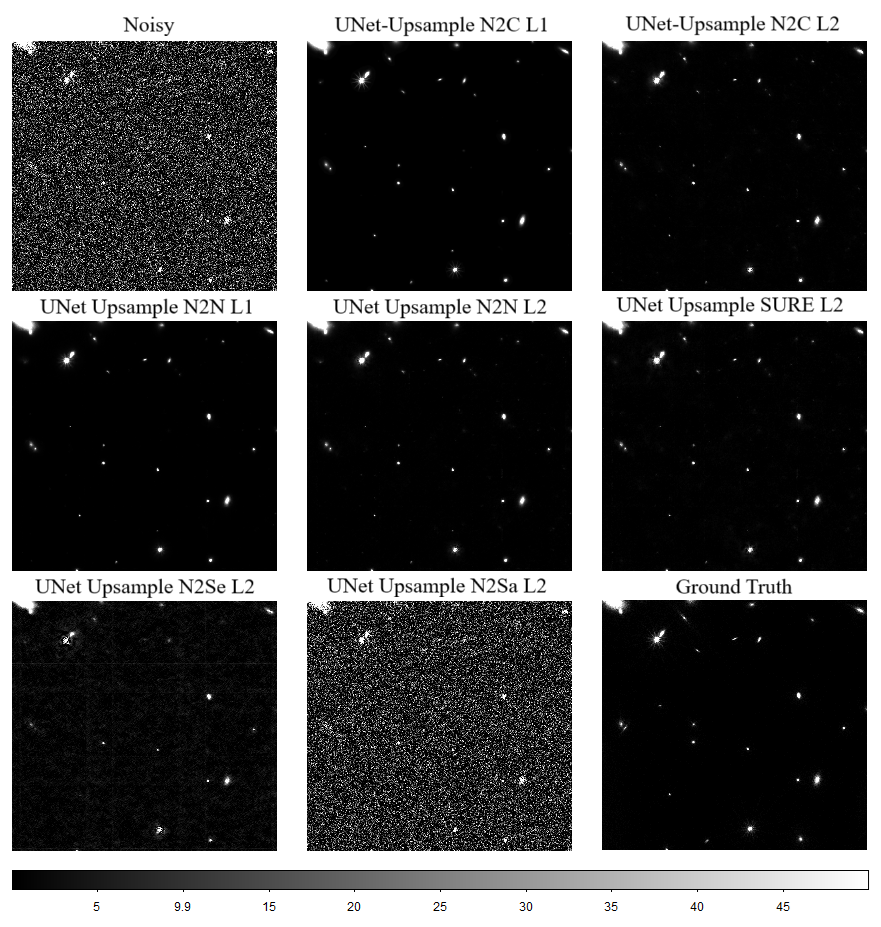}
\caption{Denoised output of denoisers given the same simulated image contaminated with the simple noise model}
\label{figure:results_simple_visualizations}
\end{figure}

The following figures provide a qualitative comparison between the outputs of different denoisers when applied to the same synthetic image. \autoref{figure:results_simple_visualizations} shows the denoised output of the U-Net with upsampling layers trained using different algorithms. The title of the frames shows the network architecture, the algorithm on which the model was trained, and the loss function used. The results show that the denoisers trained using the Noise2Clean, Noise2Noise, and SURE algorithms generate exceptional results, almost replicating the ground truth. While the $L_1$ loss is more successful in keeping the fine details intact, the $L_2$ loss is shown to smooth the surroundings of the sources more heavily. The Noise2Same results illustrate how the denoiser trained with this method fails to denoise the input image. The denoiser trained with the Noise2Self strategy also introduces noticeable grid-like artifacts. Although inferences and detections are conducted on full-field images, the figures presented here visualize only smaller patches of the full-field image. Even with the visualization scalars applied, displaying the entire field of view does not reveal discernible differences between the outputs.

\begin{figure}[!htb]
\centering
\includegraphics[width=0.85\linewidth]{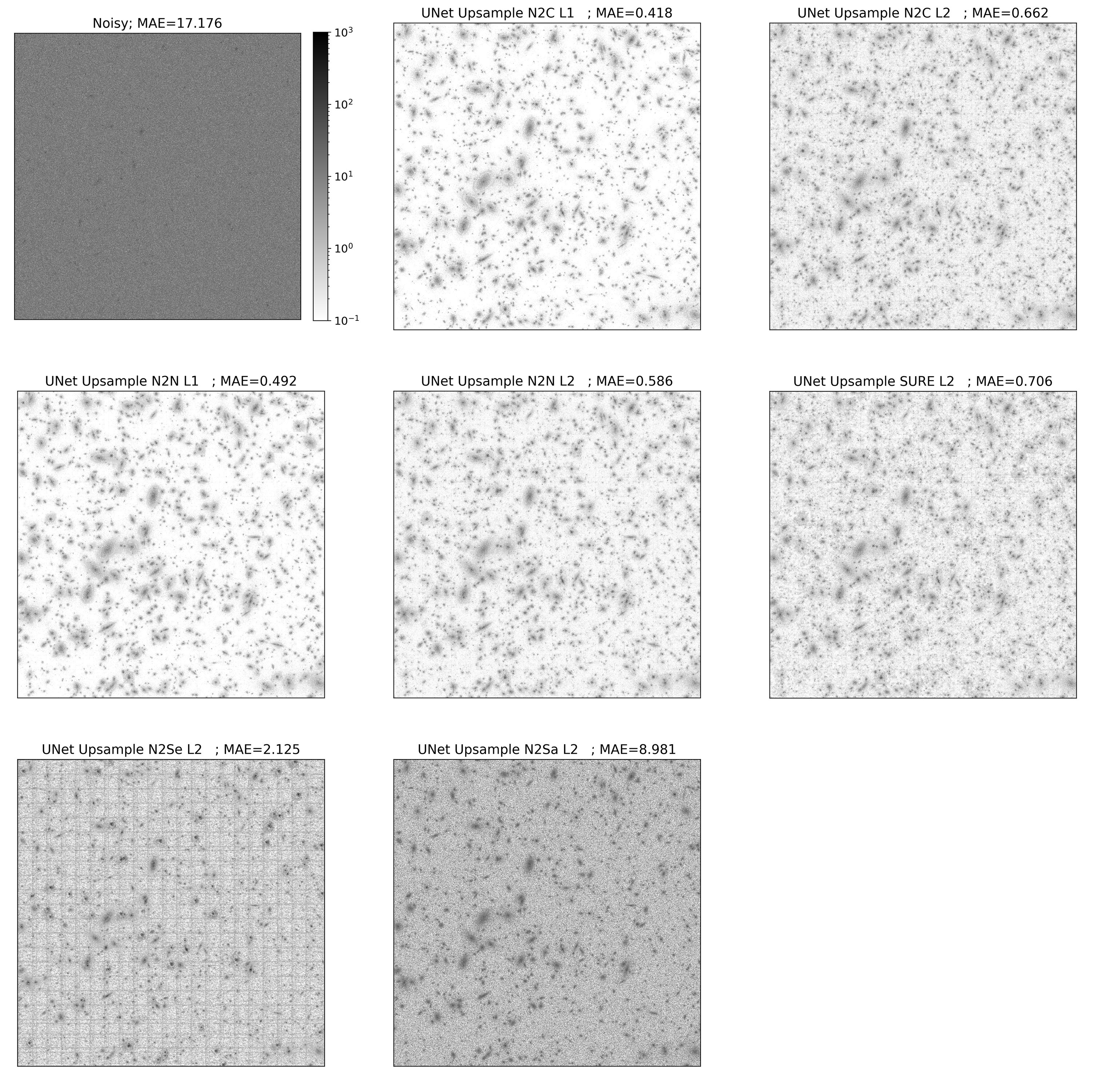}
\caption{Error maps showing the difference between the denoised outputs of each image and the ground truth target; The image is simulated and contaminated with the simple noise model.}
\label{figure:results_simple_errors}
\end{figure}

\autoref{figure:results_simple_errors} depicts the error map\footnote{Error map shows the difference between the output of a model given a noisy image and the expected noiseless image.}, of the denoised outputs of the various denoisers given the same input image. Models were given a simulated full-field image polluted with the simple noise model as input. The results of all of the Noise2Clean, Noise2Noise, and SURE denoisers show lower error levels overall compared to the input image. It can be observed that the primary source of the error originates from the pixels containing bright objects. Precisely estimating the values of these objects is nearly impossible without prior knowledge of each galaxy's morphology. Even with this minimal signal loss, raw frames are usually favored over any denoised or stacked image during subtle analysis. However, we prove a significant improvement in detection performance after denoising compared to raw images. $L_1$-based denoisers are performing marginally better at eliminating Gaussian noise compared to $L_2$-based denoisers. Noise2Same's inability to train the network causes additional artifacts added to the noisy input. The artifacts introduced by the Noise2Self method are also visible in this figure.

\subsubsection{Detailed Noise model}

In the next phase, we redo the previous experiment on synthetic images contaminated by the detailed noise model. It is a more complex problem than the sole mixture of Poisson and Gaussian noise. It is more realistic and does not match the denoising algorithms' assumptions of inter-pixel independence and zero-mean noise. However, with a few simplifying assumptions discussed in \autoref{sec:noise_formation}, the detailed noise model can be brought within the necessary conditions, including

\autoref{tab:detailed_noise} highlights the significant improvements achieved by denoisers trained under the Noise2Clean, Noise2Noise, and SURE frameworks. Although promising in theory, the Noise2Self and Noise2Same approaches underperform in this experimental setting. The denoisers trained with the $L_2$ loss perform better in the second-order error metrics, such as MSE, SNR, and PSNR, while $L_1$-based denoisers surpass them in MAE and detection performance. 

\begin{table*}[!htb]
\centering
\caption{Performance comparison of denoisers on simulated images contaminated with the detailed noise. The models are trained with simulated data and the detailed noise model.}
\label{tab:detailed_noise}
\footnotesize
\renewcommand{\arraystretch}{1.5}
\begin{tabular}{blbcccc|cc|cccb}
\bighline
Denoiser & 
MAE & 
MSE &
PSNR (dB)& 
SNR (dB) & 
KL Div. & 
NIQE & 
Detection Count & 
FAR (\%) &
CDR (\%) \\
\bighline
N2C L1     & \textbf{0.60} & 2.59 & 92.55 & 42.39 & \textbf{0.07} & 20.49 & 3039.73 & 0.16 & \textbf{81.00} \\
N2C L2     & 0.77 & \textbf{2.45} & \textbf{92.83} & \textbf{42.69} & \textbf{0.07} & 19.05 & 3017.33 & \textbf{0.21} & 80.39 \\
N2N L1     & 1.11 & 92.99 & 76.89 & 26.46 & 0.19 & 18.32 & 3004.86 & 0.09 & 80.11 \\
N2N L2     & 1.42 & 105.51 & 76.34 & 25.88 & 0.27 & 20.06 & 2978.16 & 0.14 & 79.45 \\
N2Se L2    & 2.91 & 12368.00 & 55.67 & 2.42 & 0.80  & 18.94 & 2836.61 & 0.20 & 76.29 \\
N2Sa L2    & 4.44 & 162.01 & 74.35 & 23.87 & 1.38 & 32.38 & - & - & - \\
SURE L2    & 1.65 & 130.62 & 75.42 & 24.95 & 0.46 & 19.82 & 3048.76 & 1.37 & 80.48 \\
\hline \hline
Noisy Image & 6.99 & 188.9 & 73.64 & 23.17 & 2.96 & 22.40 & 2771.00 & 2.26 & 73.05 \\
\bighline
\end{tabular}
\end{table*}

We further investigate how models trained with simple noise formation models perform on images contaminated by detailed noise. This is done to check the generalizability of approaches. \autoref{tab:detailed_noise2} shows that the Noise2Noise training scheme is performing better than the supervised approaches, suggesting better generalizability to different domains and noise models. Most of the denoisers deliver acceptable results and metrics on images affected by the detailed noise model. The denoiser trained using the SURE framework demonstrates promising improvements in the signal, while the denoisers trained with the Noise2Self and Noise2Same algorithms underperform. 

\begin{table*}[!htb]
\centering
\caption{Performance comparison of denoisers on simulated images contaminated with the detailed noise. The models are trained with simulated data and the simple noise model.}
\label{tab:detailed_noise2}
\footnotesize
\renewcommand{\arraystretch}{1.3}
\begin{tabular}{blbcccc|cc|cccb}
\bighline
Denoiser & 
MAE & 
MSE &
PSNR (dB)& 
SNR (dB) & 
KL Div. & 
NIQE & 
Detection Count & 
FAR (\%) &
CDR (\%) \\
\bighline
N2C L1  & 1.31 & 117.60 & 75.87 & 25.39 & 0.26 & 20.74 & 2974.35 & \textbf{0.06} & 79.38 \\
N2C L2  & 1.36 & 110.77 & 76.13 & 25.66 & 0.25 & 18.51 & 2881.63 & 0.12 & 77.00 \\
N2N L1  & \textbf{1.08} & 111.35 & 76.11 & 25.63 & \textbf{0.17} & 19.82 & 2979.02 & 0.07 & \textbf{79.49} \\
N2N L2  & 1.29 & \textbf{109.32} & \textbf{76.19} & \textbf{25.72} & 0.23 & 19.26 & 2885.45 & 0.08 & 77.11 \\
N2Se L2 & 4.39 & 7109.93 & 58.07 & 7.12 & 1.19 & 25.89 & 3013.98 & 13.50 & 71.65 \\
N2Sa L2 & 7.03 & 191.29 & 73.59 & 23.12 & 2.96 & 22.31 & - & - & - \\
SURE L2 & 1.32 & 121.90 & 75.71 & 25.22 & 0.24 & \textbf{16.83} & 2819.37 & 0.17 & 75.42 \\
\hline \hline
Noisy Image & 6.99 & 188.9 & 73.64 & 23.17 & 2.96 & 22.40 & 2771.00 & 2.26 & 73.05 \\
\bighline
\end{tabular}
\end{table*}

\subsection{Effects of Detection Parameters on Different Denoisers} \label{sec:detection_paraemter}

\begin{figure}[!htb]
\centering
\includegraphics[width=1\linewidth]{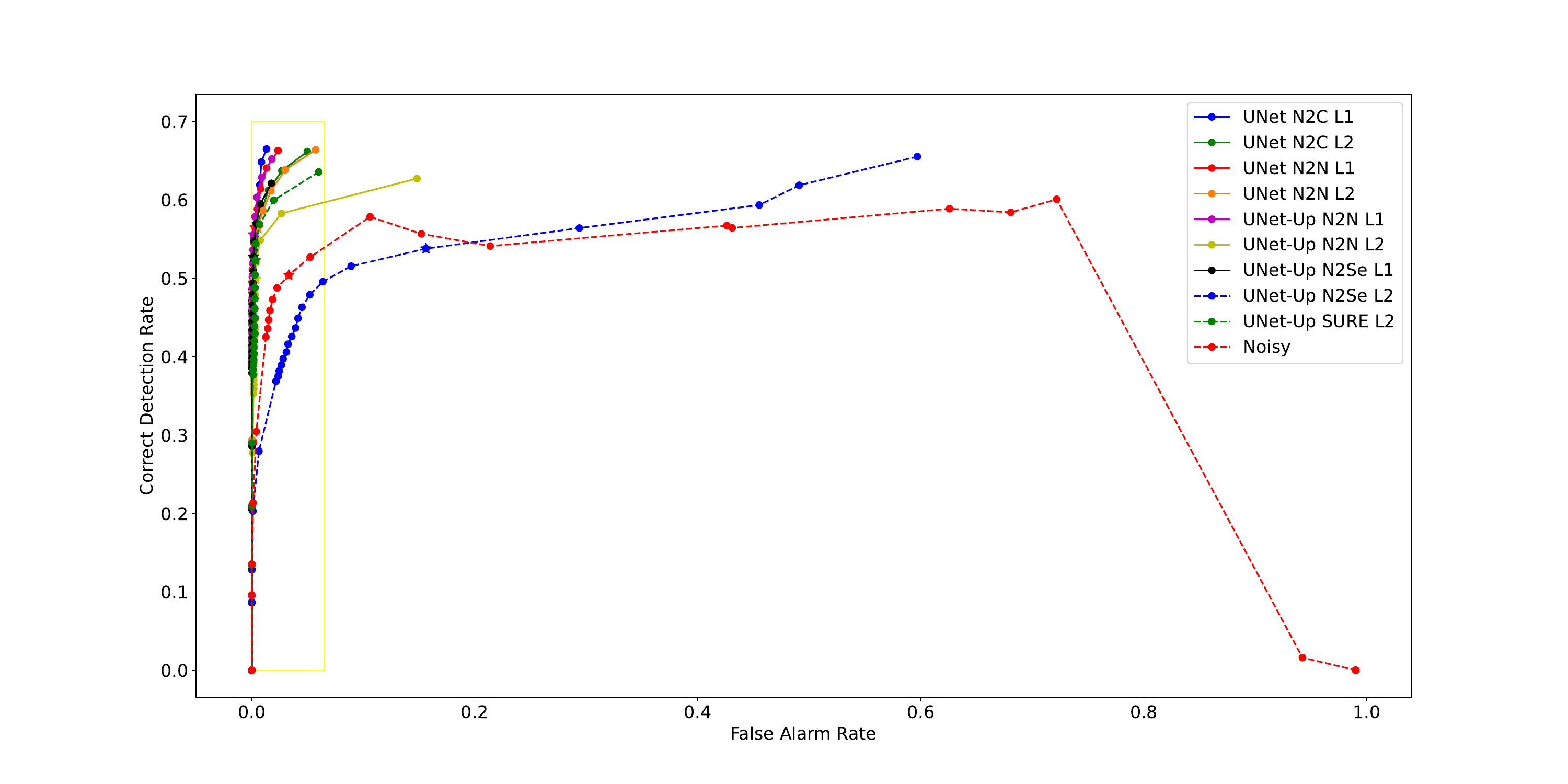}
\textbf{(a)}
\includegraphics[width=1\linewidth]{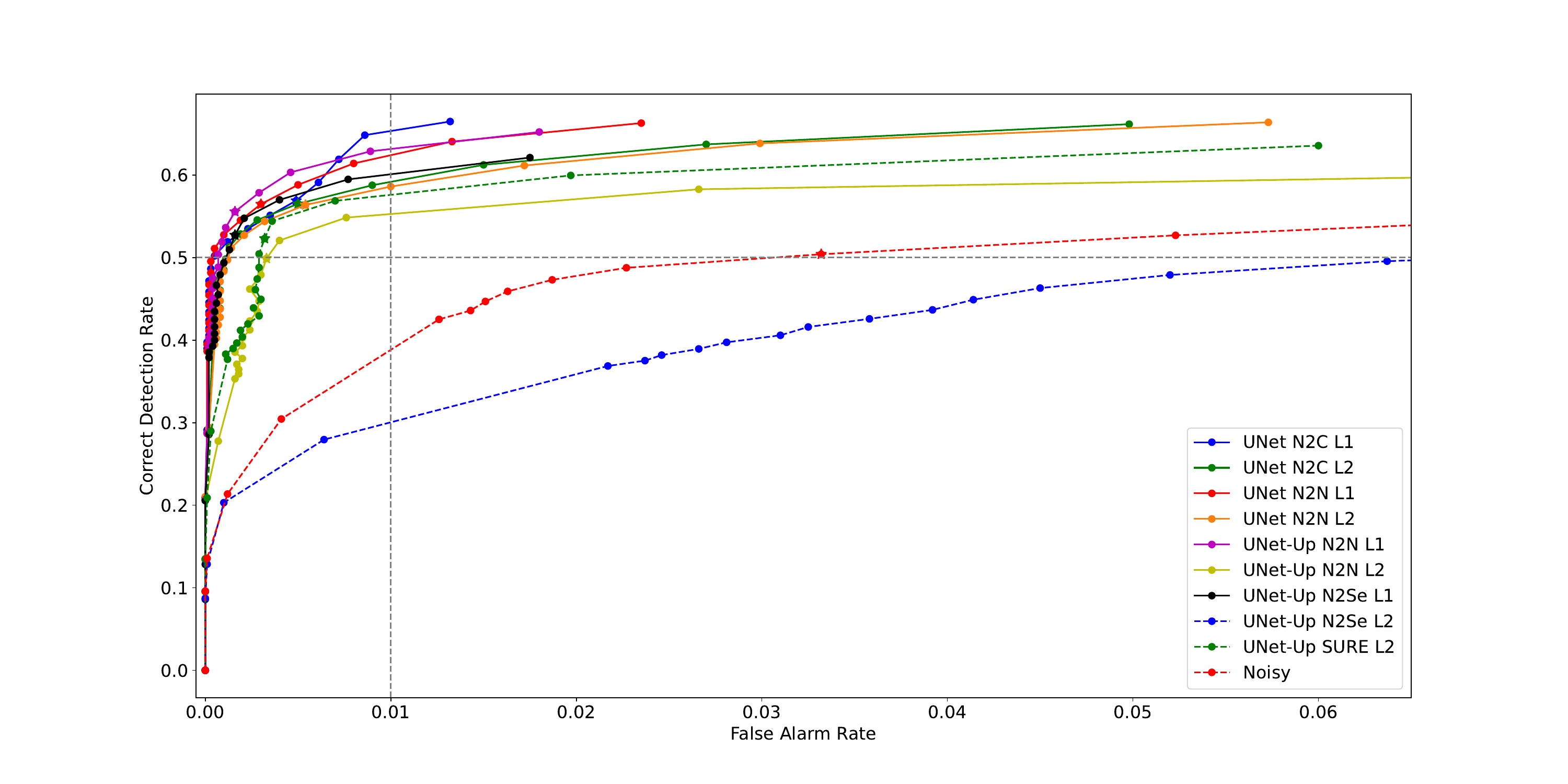}
\textbf{(b)}
\caption{Correct Detection Rate vs. False Alarm Rate comparison of different training strategies; Subfigure \textbf{(b)} shows the selected range shown in subfigure \textbf{(a)} by the yellow rectangle.}
\label{figure:roc}
\end{figure}

The final experiment done on the simulated data is to test the effect of the detection parameters. The threshold scalar used for the SExtractor's detection is $\sigma=3$ for the noisy frames and $\sigma=1.5$ for the denoised images previously shown. This parameter is multiplied by the estimate of the background root mean square (background RMS)\footnote{The background root mean square refers to the standard deviation of the pixel values in a region of an astronomical image.} at each pixel and counts all pixels above their threshold as bright source candidates. The $\sigma=1.5$ and $\sigma=3$ are two common values for this parameter. The $\sigma=3$ can be used for the denoised images instead of the $\sigma=1.5$. However, the denoising algorithm allows us to lower this threshold, improving the detection performance. We sweep the relative threshold parameter of the SExtractor algorithm from 0.5 to values up to 5000. \autoref{figure:roc} depicts the correct detection rate versus the false alarm rate analysis of different training strategies. The top panel shows the full range of the analysis, and the bottom panel is the magnified version of the yellow rectangle in the top panel. For this experiment, we use the modified U-Net on the synthetic dataset contaminated with the simplified noise model.

The average false alarm rate of detections on the noisy frame is considerably higher than the denoised images, confirming that the denoised frames offer a more balanced trade-off between correct detection and false alarm rates. The highest percentage of objects correctly detected on the noisy frames with an acceptable false alarm rate of 1\% is around 38\%, which is far less than the CDR of denoised images. Even with a false alarm rate tolerance increased to 10\%, the number of reference objects detected in the noisy images remains lower than in the denoised outputs of all denoising algorithms except Noise2Self. This occurs despite the negligible errors in the detection done on the denoised images. The points visualized by the asterisk sign are the points reported in the \autoref{tab:simple_noise}. The most stable denoiser in the detection is the modified U-Net with upsampling layers trained with the Noise2Noise and $L_1$ loss, closely followed by other Noise2Noise, Noise2Clean, and SURE denoisers.

\autoref{tab:detection_comparison}'s left table compares the performance of each denoiser when subjected to a maximum false alarm rate threshold of 1\%. The table on the right, however, examines each denoiser's minimum false alarm rate, given a minimum correct detection rate of 50\%. The results indicate that denoisers trained with $L_1$ loss using the Noise2Clean and Noise2Noise settings achieve the best performance in general, and the results of the denoiser trained with the SURE framework and Noise2Self under the $L_1$ loss are slightly lagging.

\begin{table*}[!htb]
\centering
\caption{Comparison of detection performance metrics. (a) Performance with a maximum false alarm rate (FAR) of 1\%. (b) Performance with a fixed correct detection rate (CDR) of 50\%.}
\label{tab:detection_comparison}
\footnotesize
\renewcommand{\arraystretch}{1.3}

\begin{minipage}[t]{0.48\linewidth}
    \centering
    \textbf{(a) Max FAR = 1\%} \\ \vspace{5pt}
    \begin{tabular}{|l|cc|}
    \hline
    Denoiser & FAR (\%) & CDR (\%)\\
    \hline
    U-Net N2C L1     & 0.86 & \textbf{64.84} \\
    U-Net N2C L2     & 0.90 & 58.77 \\
    U-Net N2N L1     & 0.80 & 61.41 \\
    U-Net N2N L2     & 1.00 & 58.61 \\
    U-Net-Up N2N L1  & 0.89 & 62.88 \\
    U-Net-Up N2N L2  & 0.76 & 54.85 \\
    U-Net-Up N2Se L2 & 0.64 & 27.95 \\
    U-Net-Up SURE L2 & 0.70 & 56.88 \\
    \hline \hline
    Noisy Image      & 0.41 & 30.45 \\
    \hline
    \end{tabular}
\end{minipage}%
\hfill 
\begin{minipage}[t]{0.48\linewidth}
    \centering
    \textbf{(b) Fixed CDR = 50\%} \\ \vspace{5pt}
    \begin{tabular}{|l|cc|}
    \hline
    Denoiser & FAR (\%) & CDR (\%)\\
    \hline
    U-Net N2C L1     & \textbf{0.05} & 50.21 \\
    U-Net N2C L2     & 0.13 & 51.31 \\
    U-Net N2N L1     & \textbf{0.05} & 51.10 \\
    U-Net N2N L2     & 0.14 & 51.17 \\
    U-Net-Up N2N L1  & 0.07 & 50.36 \\
    U-Net-Up N2N L2  & 0.40 & 52.08 \\
    U-Net-Up N2Se L2 & 8.90 & 51.55 \\
    U-Net-Up SURE L2 & 0.29 & 50.45 \\
    \hline \hline
    Noisy Image      & 15.21 & 55.67 \\
    \hline
    \end{tabular}
\end{minipage}
\end{table*}

\subsection{Results on Real Data} \label{sec:real_data}

We next evaluate the AstroSURE workflow on two real observational datasets, one from HST and one from CFHT. In contrast to the synthetic experiments, no clean target is available for these data. We therefore rely primarily on catalogue-based detection metrics, together with auxiliary no-reference image-quality measures such as uPSNR and NIQE. For both datasets, we use the modified U-Net initialized from Noise2Noise pre-training on the synthetic space-based dataset and then adapt the model to the target data using the SURE loss. These experiments are intended to assess both practical detection performance and the extent to which a model trained on simulated space-based data transfers to real space- and ground-based observations.

\textbf{Hubble Space Telescope:} 

The HST WFC3 dataset is the closer of the two real-data domains to the synthetic training set, since both correspond to space-based optical imaging without atmospheric seeing. \autoref{tab:real_data_results} shows encouraging improvements after adaptation with SURE: uPSNR increases, NIQE improves, and the Correct Detection Rate rises from $31.12\%$ to $35.72\%$, while the False Alarm Rate decreases modestly by 2\%. Taken together, these results indicate that the denoised images can improve faint-source detection in this setting. They also suggest that initialization from a physically similar simulated domain is beneficial for unsupervised adaptation to real space-based data.

\textbf{Canada-France-Hawaii Telescope:} In contrast, the gains for the CFHT MegaCam data are limited. This is consistent with the larger domain gap between the seeing-limited, ground-based CFHT images and the space-based synthetic data used for initialization. Ground based telescopes are subject to atmospheric seeing, higher sky backgrounds, and different noise characteristics, compared to the Roman-like simulations used to train AstroSURE. We applied the same pipeline and initialed the network using the space-based simulation weights. In terms of image quality, \autoref{tab:real_data_results} shows a negligible improvement in uPSNR, although the increase in NIQE is akin to the HST data. The CDR increases only fractionally while the FAR of the denoised image increases slightly. This result highlights a natural limitation of the proposed method.  The initialization domain must share fundamental characteristics with the target domain, avoiding the need to generalize to conditions not experienced during training. Here, the pre-trained features learned from sharp, diffraction-limited space simulations do not transfer effectively to seeing-limited ground-based data. More extensive domain-specific retraining is required. This underscores our recommendation that for optimal performance, the unsupervised AstroSURE pipeline should be initialized with weights derived from a similar observing domain.

\begin{table}[!htb]
\centering
\caption{Quantitative evaluation of denoising and detection performance on real observational data. The image quality metrics (uPSNR and NIQE) do not require a noise-free ground truth.  The detection metrics (CDR and FAR) are computed using deep reference catalogs.}
\label{tab:real_data_results}
\small
\renewcommand{\arraystretch}{1.3}
\begin{tabular}{bl|c|cc|ccb}
\bighline
\textbf{Dataset} & \textbf{Image State} & \textbf{uPSNR (dB)} & \textbf{NIQE} & \textbf{CDR (\%)} & \textbf{FAR (\%)} \\
\bighline
\multirow{2}{*}{\textbf{HST (Space)}} & Noisy Input & 43.01 & 22.70 & 31.12 & 66.76 \\
 & Denoised (Ours) & \textbf{50.72} & \textbf{16.55} & \textbf{35.72} & \textbf{65.49}  \\
\hline
\multirow{2}{*}{\textbf{CFHT (Ground)}} & Noisy Input & 40.14 & 21.94 & 20.29 & \textbf{44.74}\\
 & Denoised (Ours) & \textbf{40.48} & \textbf{16.55} & \textit{20.80} & 46.24\\
\bighline
\end{tabular}
\end{table}

\section{Conclusion and Future Works}\label{sec:conclusion}
In conclusion, this study examines the training strategies used for natural-scene images and evaluates the applicability of these approaches to astronomical images. We introduce the AstroSURE training pipeline to denoise astronomical images without access to ground truth. Whenever multiple noisy frames from the same field of view are available for training, it can use Noise2Noise, which performed best among the methods tested here. For scenarios where only single noisy frames of each field are available, we propose using the SURE framework for training.  SURE can achieve results comparable to those of Noise2Noise.

We demonstrate improvements in signal metrics and detection on both synthetic and real data.  Ultimately, these techniques will reduce the number of exposures required to acquire a strong signal. Unlike existing training methods with clean and noisy datasets, our approach achieves comparable results using only noisy exposures without relying on stacked images. This framework exhibits potential for broader applicability to diverse astronomical images. The proposed solution can be employed in any architecture, paving the way for future adaptations utilizing larger, state-of-the-art Vision Transformer (ViT) models to better capture complex spatial dependencies.

When testing AstroSURE on real-world HST and CFHT data, we observed that the imaging conditions used to synthesize the pre-training data must closely align with the target real-world data. Specifically, our results demonstrate stronger performance on the space-based HST data compared to the ground-based CFHT data. Consequently, an immediate direction for future research is to expand our experiments to encompass extensive datasets from a wider variety of telescopes, with a specific focus on mitigating the unique instrumental artifacts inherent to ground-based observatories.

Alongside expanding dataset diversity, another critical avenue for future work is the treatment of structured noise. Although our current study does not address these specific artifacts, our approach demonstrates strong potential for direct integration into raw astronomical imaging pipelines. Moving forward, subsequent experiments will be essential to determine how structured contaminants interfere with the proposed denoisers and to develop strategies for mitigating these effects during training.

To address this upcoming challenge, the reconstruction of images corrupted by structured contaminants can be modeled as an image inpainting problem. For instance, methods such as Robust Equivariant Imaging \footnote{Robust Equivariant Imaging extends denoising losses like Noise2Noise and SURE into a training framework called equivariant imaging, which is designed to handle structured contaminants.} can be adapted for most artifacts where the noise function remains static throughout the dataset \citep{rei}. Conversely, for dynamic contamination sources like cosmic rays, which affect completely different sets of pixels across different frames, future implementations could leverage extended approaches, such as those proposed by \citet{rei_extended}, to solve inpainting problems with varying measurement operators.


\par\bigskip
\par\bigskip
\noindent{\bf Acknowledgments:}

The authors acknowledge the use of the Canadian Advanced Network for Astronomy Research (CANFAR) Science Platform. Our work used the facilities of the Canadian Astronomy Data Center, operated by the National Research Council of Canada with the support of the Canadian Space Agency, and CANFAR, a consortium that serves the data-intensive storage, access, and processing needs of university groups and centers engaged in astronomy research \cite{canfar}.

This research was enabled in part by support provided by \href{https://www.computeontario.ca/}{Compute Ontario} and the  \href{alliancecan.ca}{Digital Research Alliance of Canada}.

This work has made use of data from the European Space Agency (ESA) mission
\href{https://www.cosmos.esa.int/gaia}{\it Gaia}, processed by the {\it Gaia} Data Processing and Analysis Consortium (\href{https://www.cosmos.esa.int/web/gaia/dpac/consortium}{DPAC}). Funding for the DPAC has been provided by national institutions, in particular, the institutions participating in the {\it Gaia} Multilateral Agreement.

%



\software{Astropy \citep{2013A&A...558A..33A,2018AJ....156..123A},
          Galsim \citep{galsim}
          Maximask \cite{maximask}, 
          Source Extractor \citep{1996A&AS..117..393B},
          Topcat \& still \citep{stilts}
          }


\newpage
\appendix

\section{Noise Formation Model} \label{sec:noise_formation}
\begin{figure}[H]
\centering
\includegraphics[width=0.6\linewidth]{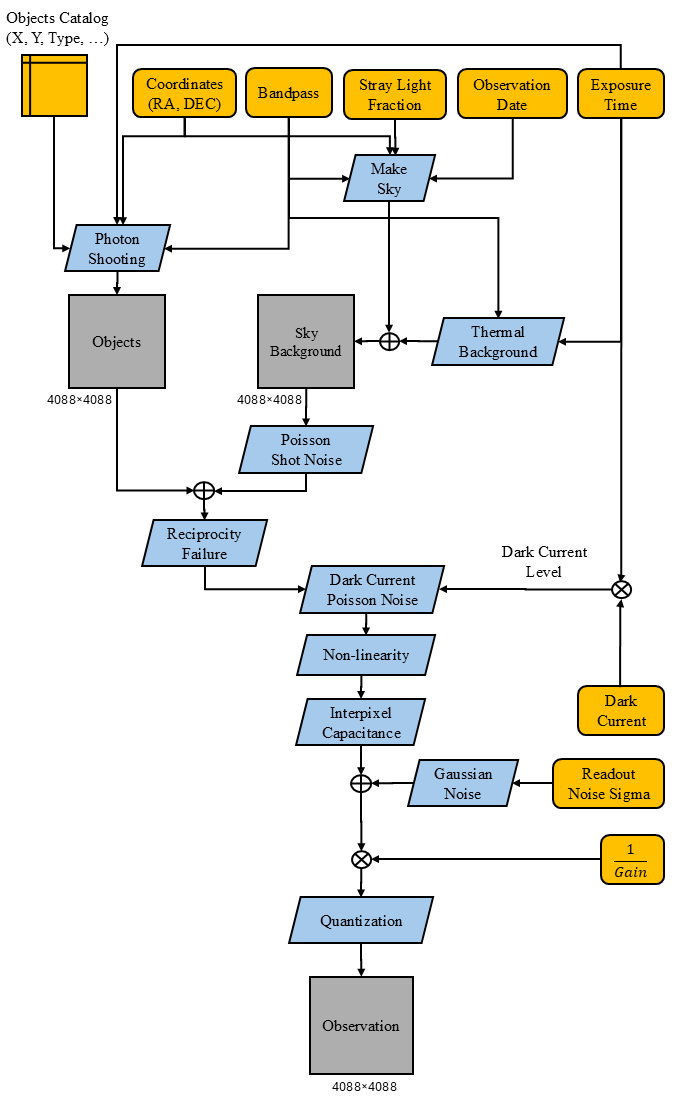}
\caption{Overview of the noise formation model synthesized with Galsim}
\label{figure:noise_formation}
\end{figure}

\section{Impact of Pre-processing} \label{sec:scalings}

In this experiment, we examine how different scaling methods of the range of the data and different preprocessing affect the training process. \autoref{tab:scaling} illustrates that keeping the data in the range of $0$ and $65536$ yields better results compared to the following scaling approaches: (I) dividing the pixel values by the maximum pixel value of $65536$, (II) normalizing between $0$ and $1$, (III) standardizing to zero mean and variance of 1, and (IV) inverse hyperbolic sine function. Also, clipping the denoised output results in more stable training and better results. The normalization scaling allows easier generalizability of the trained models and allows them to be applied to datasets with different ranges. However, the best-performing pre-processing was no scaling with clipping. This result further suggests that pre-processing choices and domain mismatch can materially affect denoiser performance when transferring a model to a new instrument.

\begin{table*}[!htb]
\centering
\caption{Comparison of scaling and clipping for denoising}
\label{tab:scaling}
\renewcommand{\arraystretch}{1.3}
\begin{tabular}{|l|ccc|c|}
\bighline
Preprocessing & MAE & MSE & PSNR (dB) & KL Div. \\
\hline
No Scaling with Clipping        & \textbf{1.02} & \textbf{20.24} & \textbf{91.45} &\textbf{ 2.79} \\
No Scaling without Clipping     & 1.93 & 28.84 & 90.35 & 6.26 \\
Division                        & \textbf{1.02} & 25.53 & 85.35 & 6.45 \\
Normalization                   & 1.16 & 55.95 & 87.41 & 10.16 \\
Standardization                 & 1.56 & 30.32 & 86.15 & 3.46 \\
$\rm {Sinh}^{-1}$                   & 1.58 & 754.26 & 85.65 & 5.24 \\
\hline \hline
Noisy Image & 21.26 & 836.91 & 68.39 & 16.40 \\
\bighline
\end{tabular}
\end{table*}


\section{Impact of Denoising Algorithm on the Objects Detected}
\begin{figure}[H]
\centering
\includegraphics[width=1\linewidth]{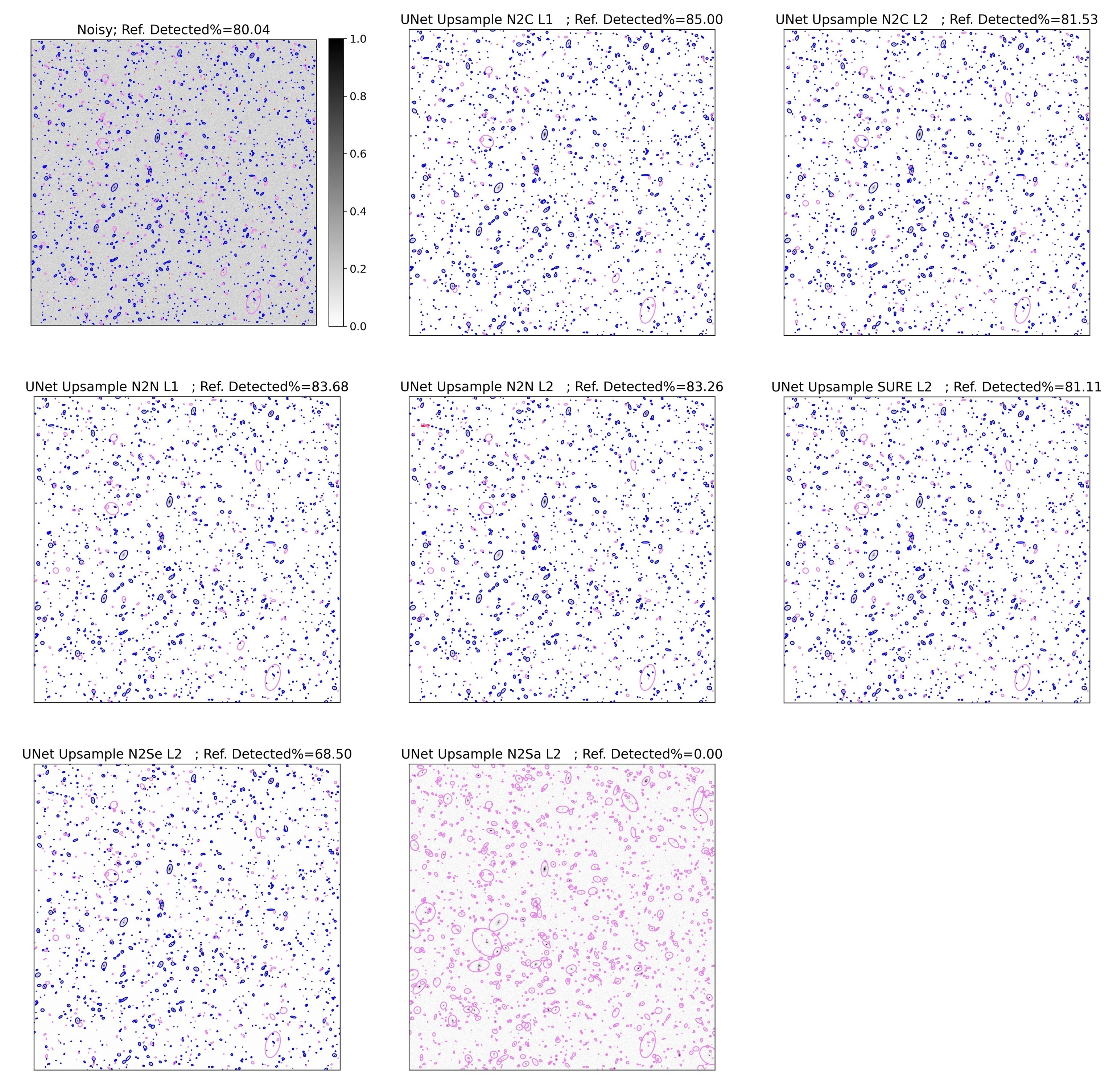}
\caption{Comparison of different denoisers in object detection on a simulated image contaminated with the simple noise model. Blue ellipses show the detections that were successfully associated with a reference object. The pink ellipses visualize the objects missed in the detection. The red ellipses, mainly visible in the noisy frame, show the false alarm detection, which the cross-matching algorithm could not associate with any reference object.}
\label{figure:results_simple_objects}
\end{figure}

\autoref{figure:results_simple_objects} illustrates the improvements in the detection of the denoised images compared to the noisy input. The tiles of the frames show similar information to the previous figures, but in this figure, the percentage of the total objects in the field that are correctly detected from the denoised outputs is shown. The blue and red ellipses represent correct detections and false alarms. The size, angle, and shape of these ellipses represent the relative dimensions of the detections (expanded by a factor of 6 for better visualization). The missed objects shown in pink are drawn based on the known object dimensions. The detection algorithm is able to detect 80.04\% of objects in the noisy frame; however, multiple red ellipses are shown scattered through the image. The small size of these false detections shows that only a few pixels were detected by the SExtractor algorithm as source objects due to noise pollution. The denoised outputs from Noise2Clean, Noise2Noise, and SURE show fewer visible false detections than the noisy frame, together with higher detection rates. The detection rate is also higher for these frames. The results show that faint objects are detected more easily in these images, while low surface brightness objects are still challenging to detect. The center bottom panel, which is the detection results of the Noise2Same, shows no correct detection. The model is disrupting the background and signal, causing the detection algorithm to fail. The SExtractor detection algorithm has a maximum threshold for the number of objects and source pixels above the background, and when this threshold is exceeded in a problematic image, faulty behaviors happen, such as no objects being detected or thousands of detections occurring in a small patch. The results of the Noise2Self denoisers show the lower detection performance of the denoised output compared to the noisy frame.

\section{Training Loss of Different Training Strategies}
\begin{figure}[!htbp]
\centering
\subfigure[Noise2Clean]{\includegraphics[width=0.49\textwidth]{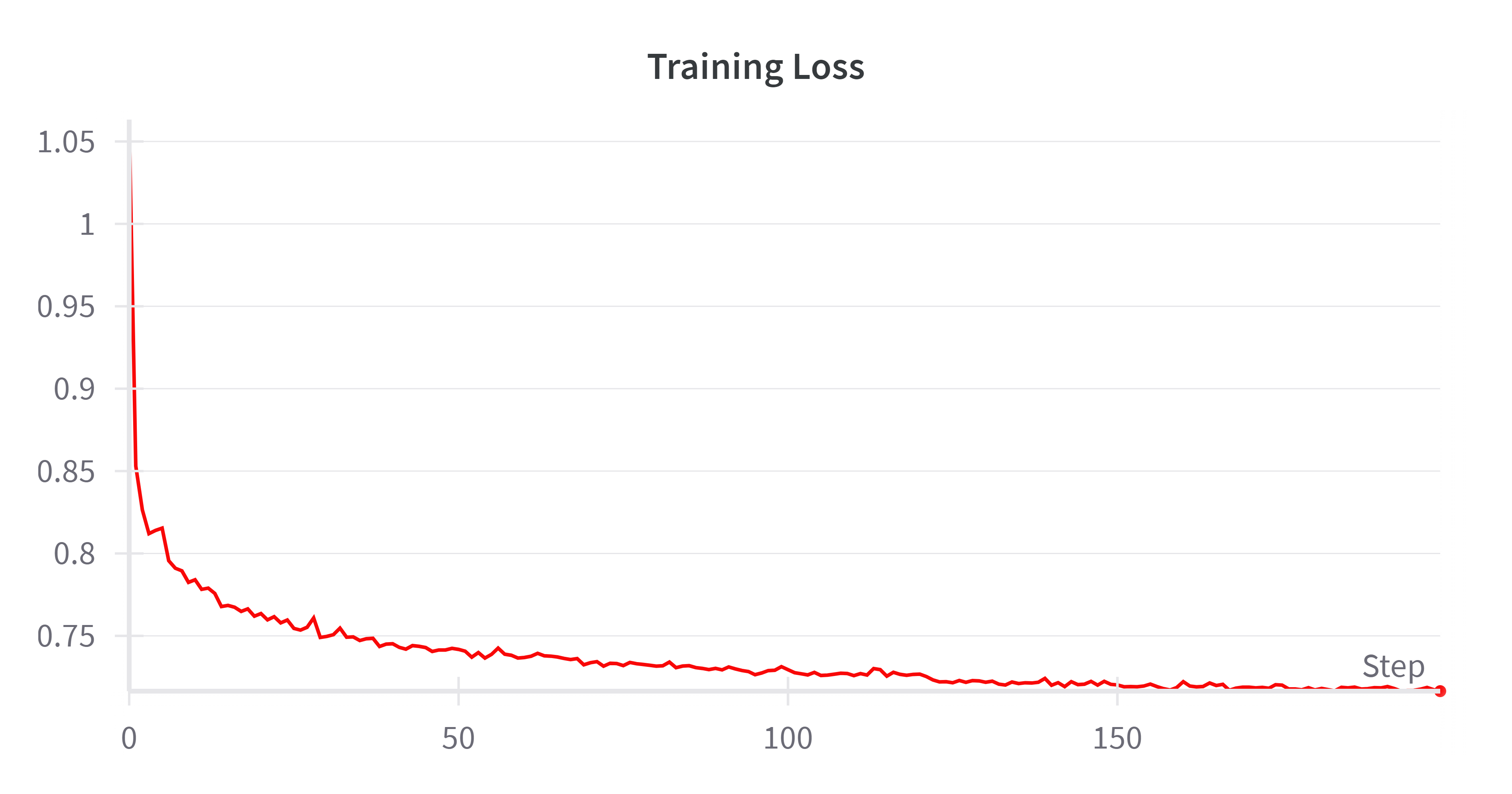}}
\subfigure[Noise2Noise]{\includegraphics[width=0.49\textwidth]{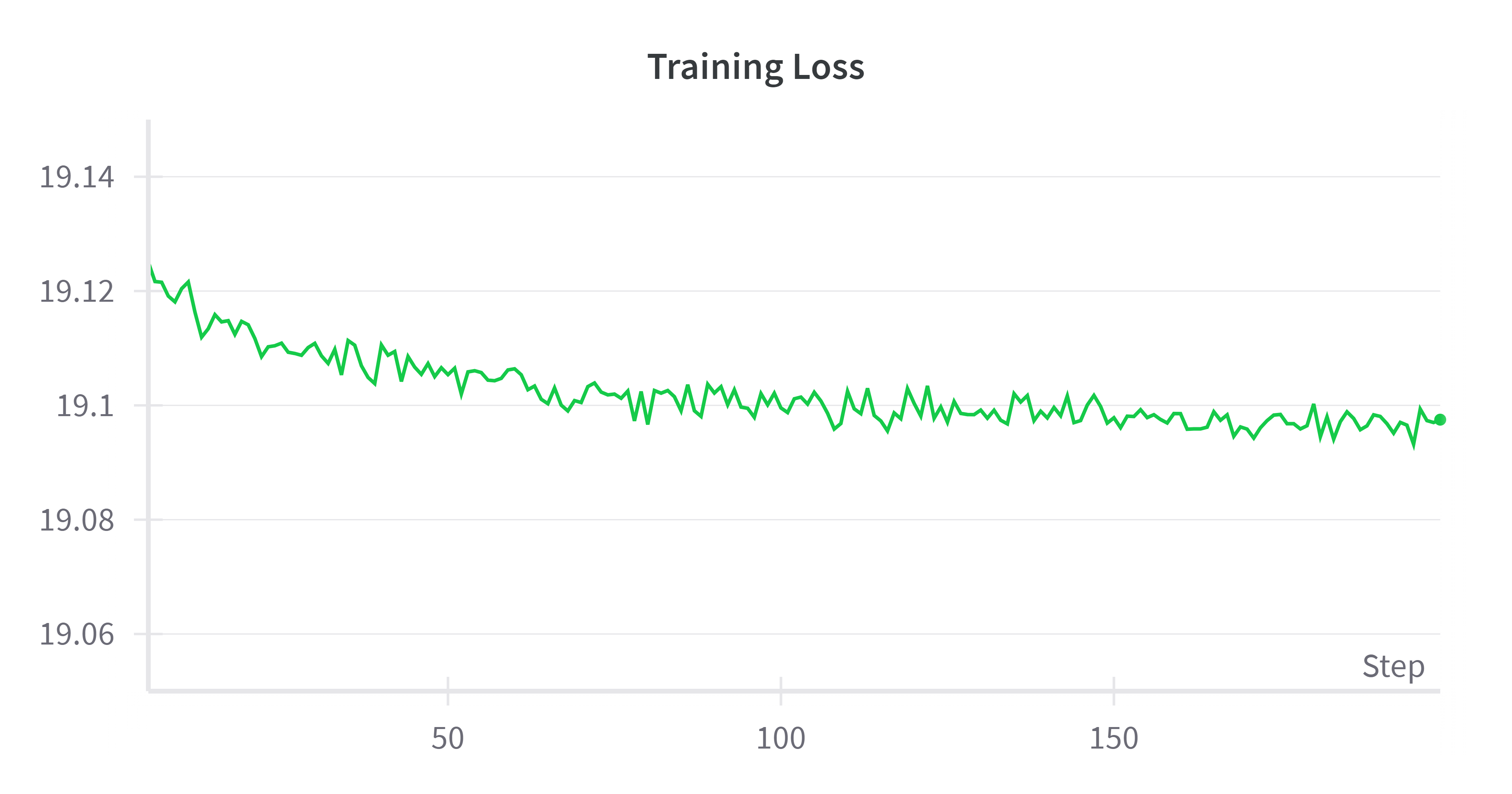}}

\subfigure[Noise2Self]{\includegraphics[width=0.49\textwidth]{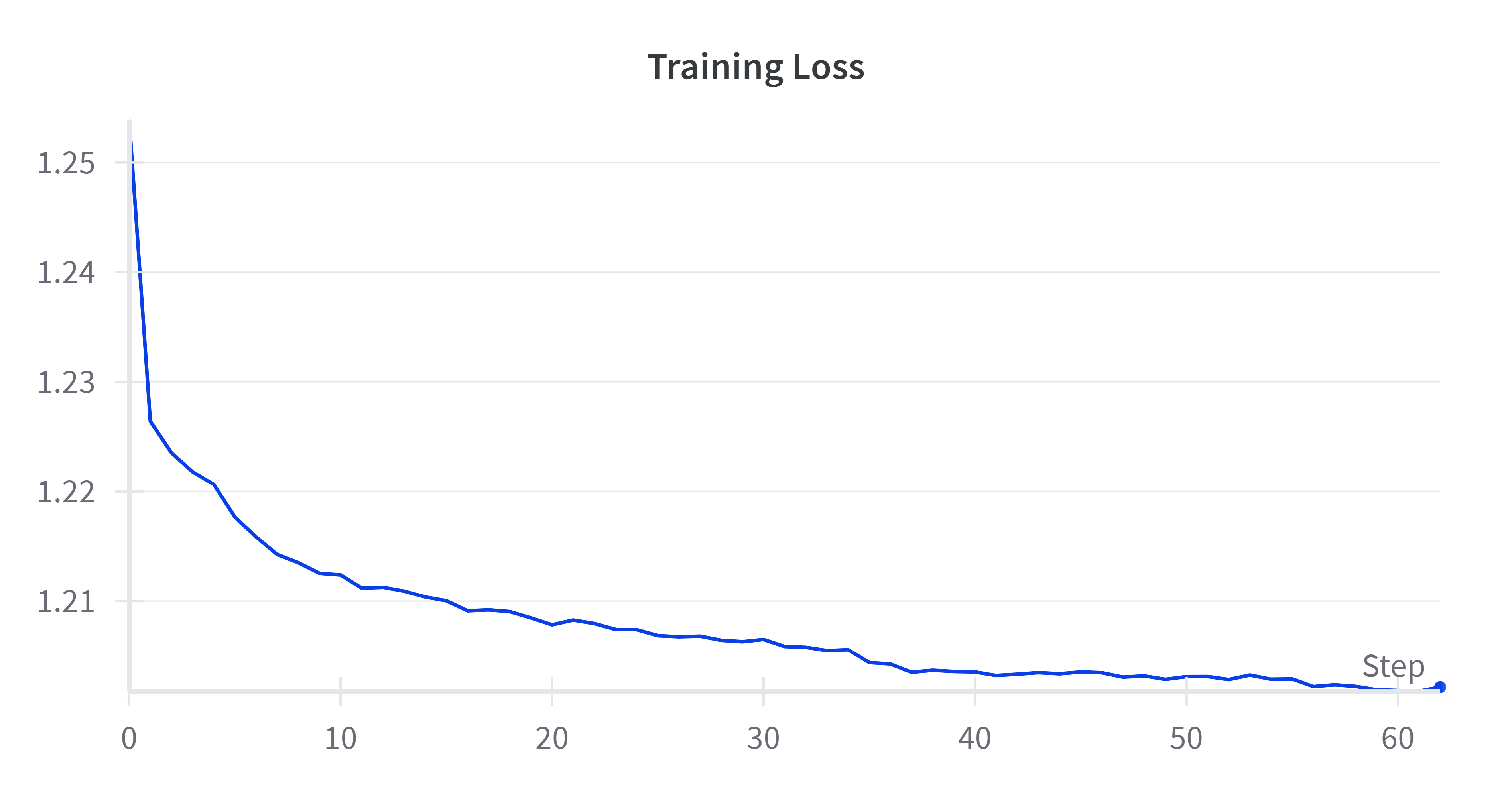}}
\subfigure[SURE]{\includegraphics[width=0.49\textwidth]{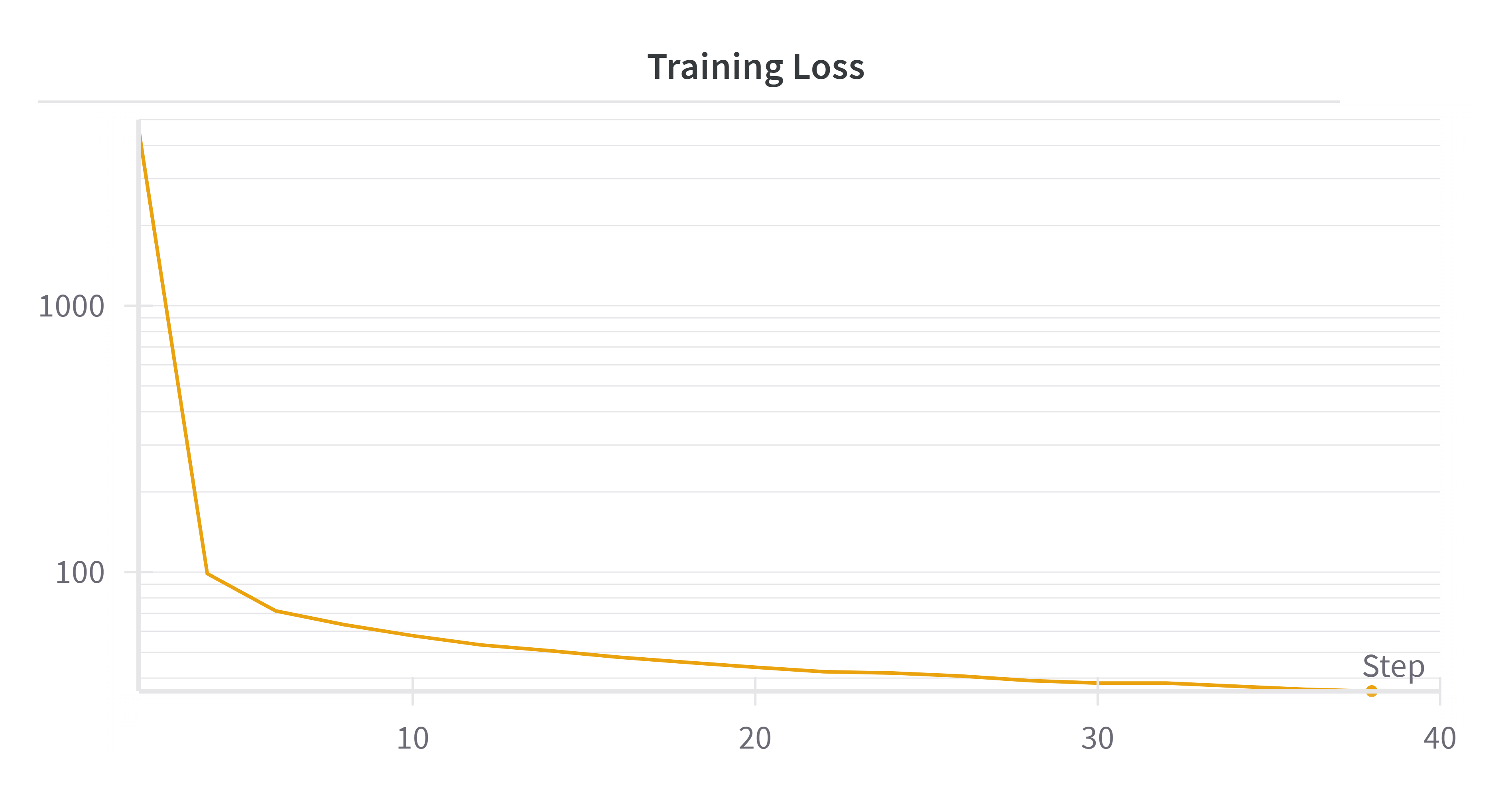}}

\caption{Training loss curve during training of the modified U-Net using different training settings; (a), (b), and (c) depict the $L_1$ training loss used during training. (d) shows the $L_2$ training loss on a logarithmic scale during training}
\label{figure:training_comparison}
\end{figure}

\autoref{figure:training_comparison} visualizes the training loss of the modified U-Net trained using Noise2Clean, Noise2Noise, Noise2Self, and SURE frameworks. It is important to note that panels (a), (b), and (c) present the $L_1$ training loss, while panel (d) shows the SURE framework using $L_2$ loss. During Noise2Clean training, the loss decreases consistently, whereas the training loss for Noise2Noise remains elevated. This can be attributed to the inherent challenge of Noise2Noise training, which involves predicting random noise, which is an impossible task, resulting in minimal improvement in training loss. The improvements in Noise2Self are also marginal. However, the SURE loss begins at high values and decreases significantly within the first few epochs. All methods are allowed a maximum of 200 training epochs, but training is terminated early if the validation loss stops improving. Hence, the number of epochs used to train each algorithm differs.

\bibliography{refs}{}
\bibliographystyle{aasjournal}



\end{document}